\documentclass[
aps,
nofootinbib,
superscriptaddress,
tightenlines,
notitlepage,
twocolumn,
showpacs,
floatfix,
]{revtex4-1}
\usepackage{amssymb,amsmath,bm,tensor,braket}
\usepackage[colorlinks]{hyperref}

\usepackage{graphicx}
\usepackage{epstopdf}
\graphicspath{{figures/}}
\usepackage{dcolumn}
\usepackage{bm}
\usepackage{xcolor}

\newcommand{\BIT}{\affiliation{School of Physics, Beijing Institute of Technology, Beijing, 100081, China}}

\begin{document}

\preprint{APS/123-QED}

\title{Observational Signatures of Dyonic Black Holes in Non-Linear Electrodynamics}

\author{Sobhan Kazempour}\email{sobhan.kazempour1989@gmail.com}\BIT
\author{Sichun Sun}\email{sichunssun@gmail.com}\BIT
\author{Zhiqing He}\email{Zhi-qing\_He@outlook.com}\BIT
\author{Chengye Yu}\email{chengyeyu1@hotmail.com}\BIT


\begin{abstract}
We study the observational features of thin-disk accretion and emission spectra for static, spherically symmetric dyonic black holes in non-linear electrodynamics (NLED). In contrast to classical Maxwell theory, where electromagnetic duality assumes equal charges, the coupling parameters in NLED introduce a new gravitational potential and modify the geometry in the strong-field regime. We also study the kinematic characteristics of particles on circular orbits, such as their angular velocity, specific energy, and specific angular momentum. It is found that modifications of NLED lead to a shrinkage of the ISCO radius, resulting in a shift of the radiation spectrum thermal maximum towards smaller radii and, consequently, an increase in the local effective temperature in comparison with the standard Schwarzschild and Reissner-Nordstr\"om solutions. Additionally, we compute null-geodesic orbits to determine the optical edge of both the photon sphere and the black hole shadow. In order to assess the observational validity of the aforementioned alternatives, we carry out a $\chi^2$ goodness-of-fit test through the angular diameter constraints at the horizon scale collected by the Event Horizon Telescope (EHT) concerning Sagittarius A* ($\text{Sgr A}^*$). We prove that both these dyonic NLED and purely magnetic models have good statistical agreement with experimental data.
\end{abstract}


\maketitle


\section{Introduction}\label{sec:1}

The theory of General Relativity (GR), formulated by Einstein, has undergone over a century of rigorous observational tests, cementing its status as the leading description of gravity and spacetime geometry~\cite{Will:2014kxa}. Some of the theory's most intriguing aspects include black holes as perfect macroscopic laboratories for strong gravity. The historical progression was initiated with the discovery of the uncharged Schwarzschild solution~\cite{Schwarzschild:1916uq} and further continued with the charged Reissner-Nordstr\"om solution~\cite{Reissner:1916cle} and the rotating Kerr metric~\cite{Kerr:1963ud}. Efforts to generalize these ideal cases to include more realistic matter fields and electromagnetic corrections continue. In particular, classical linear Maxwellian electrodynamics, which works perfectly well for weak and moderate fields, is not able to take into account high-energy self-interactions of electromagnetic fields, which lead to substantial modifications of the metric structure, geodesics, and even horizon properties~\cite{Plebanski:1970zz, Salazar:1987ap, Novello:1999pg}.

Due to these theoretical limitations, non-linear electrodynamics (NLED) emerged as a compelling framework to regularize electromagnetic fields and account for high-energy quantum corrections. Historically, Born and Infeld were the first to present the NLED theory in order to avoid the infinite self-energy of point charges \cite{Born:1934gh}. Moreover, QED showed that vacuum polarization produces non-linear interactions at high energy scales, as described in the formalism of Euler and Heisenberg \cite{Heisenberg:1936nmg}. In the history of NLED studies, there have been some ups and downs; in fact, it was a purely theoretical construction until it received a new life in the context of string theory effective actions \cite{Fradkin:1985qd}. Modern gravitation theory has seen a revival of interest in NLED because it allows the construction of regular black holes and modifies the horizon structure and thermodynamics of compact objects \cite{Croney:2025cue}.

One interesting result of NLED is dyonic black holes, which carry electric and magnetic charges simultaneously. In standard Maxwell theory, for instance, electromagnetic duality implies a symmetric treatment of the two charges in the metric. However, with non-linear corrections, the coupling makes the treatment asymmetric between the two charges. Therefore, a black hole carrying a magnetic charge causes an appreciably different curvature of space-time than a similar mass carrying an electric charge. A lot of studies have been done on non-linearly charged black holes on their quasinormal modes~\cite{Toshmatov:2018ell}, weak deflection angles, and gravitational lensing effects~\cite{Eiroa:2005ag, Allahyari:2019jqz}. Moreover, the surroundings of such an object have been studied in the form of matter inflow models. Accretion studies provide a valuable diagnostic for testing NLED theories against astronomical observations.

It should be mentioned that the presence of the non-linear terms in the action violates the electric-magnetic duality present in Maxwell theory. In the linear case, the metric depends only on the sum $q^{2} + p^{2}$, and, thus, electric and magnetic charges are equivalent. After including the corrections to the NLED, such symmetry is lost. The extra terms in the field equations correspond to the effective current, $\nabla_{\mu} F^{\mu\nu} = \mu_{0} J^{\nu}$, and, therefore, the $p \leftrightarrow q$ transformation is spoiled. At the metric level, the magnetic charge $p$ is involved explicitly in the $p^{2}/r^{2}$ and $-2ap^{4}/(5r^{6})$ terms, while the electric charge $q$ is present implicitly in the cubic equation for $F_{rt}$ and the integral in Eq. (\ref{Eq2}). It is worth noting that the condition $p=0$ does not simplify the metric as much as the condition $q=0$. Thus, the asymmetry in treatment of the two charges is obvious. Nevertheless, the duality $p \leftrightarrow q$ is present at the metric and thermodynamic levels at the first perturbative order in $a,b$, and the violation of the symmetry takes place at higher orders. This issue is considered in detail in Ref. \cite{Croney:2025cue}.

It is also worth noting relevant literature on aspects of dyonic and NLED black holes. Progress has been made on different fronts regarding dyonic and NLED black holes. In Ref.\cite{Hassaine:2007py}, Hassaine and Martinez have introduced higher-dimensional electric black holes with conformally invariant NLED, which was further studied in Ref.\cite{Kokoska:2021lrn}, where Kokoska and Ortaggio managed to find the whole family of Robinson-Trautman black holes consisting of static dyonic black holes and time-dependent radiative black holes. Ramirez-Valdez et al. \cite{Ramirez-Valdez:2023gkh} proved that dyonic black holes need two potential representations since the Dirac string is an artifact of asymmetry in representation. Jeong et al.\ \cite{Jeong:2023lkc} used the doubly holographic method to calculate the Page curve for dyonic black holes and established unitary evolution through the existence of islands. In addition, Tsuda et al.\ \cite{Tsuda:2023tcs} derived the necessary conditions for nonsingular solutions of dyons and showed that Lagrangians $\mathcal{L}(\mathcal{F}, \mathcal{G})$ with two parameters can meet these conditions, whereas one-parameter Lagrangians cannot. In contrast, the thermodynamic topology of dyonic AdS black holes depends on the ensemble \cite{Gogoi:2023xzy}, and Sucu et al.\ \cite{Sucu:2025dco} studied Hawking radiation, gravitational lensing in plasma, and quantum corrections to the thermodynamics of dyonic ModMax black holes.

Spacetime observations in a strong-field regime rely on the tight coupling between particle dynamics and the radiative transfer model. The investigations of the conditions of stability of equatorial geodesics in terms of compact objects led to the establishment of the mathematical framework necessary for describing the dynamics of the matter and the ISCO both in standard and deformed metrics~\cite{Synge:1966okc, Bardeen:1972fi}. Later, Novikov, Page, and Thorne~\cite {Novikov:1973kta, Page:1974he} generalized the full-fledged relativistic hydrodynamical model of Shakura and Sunyaev~\cite{Shakura:1972te} to arbitrary covariant spacetimes. The steady-state thin accretion disk model developed by those researchers became the diagnostic tool afterward, as any modification of the gravitational potential leads to the alteration of the spectrum continuum, the emissivity, and the efficiency of co-rotating matter~\cite{Thorne:1974ve, Bambi:2012tg}. Along with the thermodynamic signature of modifications to the spacetime metric, the geometry of null geodesics determines the form of the photon sphere and the black hole shadow, thus offering an optical test of gravity \cite{Falcke:1999pj, Johannsen:2010bi}. Given the unique opportunity of direct observation of the black hole shadow of the supermassive black hole Sgr~A$^*$ by the EHT collaboration~\cite{EventHorizonTelescope:2022xqj, EventHorizonTelescope:2022wkp}, the optical and accretion signatures of the above-mentioned theoretical predictions have become real tests of gravity. Thus, studying how modifications to the Einstein field equations affect thermal signatures and black hole shadows can serve as an empirical test of GR extensions. In addition, similar studies have been performed recently \cite{Olmo:2023lil,Rosa:2023hfm,Lowell:2023kyu,Hopkins:2023vjh,Avara:2023ztw,Kobialko:2025sls,Battista:2026nsx,Gyulchev:2026kvp}.

The key motivation and objective of this investigation is to study the thin disk surrounding dyonic black holes in the context of NLED as an experimental high-energy physics tool. In other words, our goal is to achieve quantifiable observational tests of the NLED framework and gain more detailed insights about high-energy astrophysics. The NLED solution provides a new gravitational potential that affects the ISCO position. These changes, in turn, affect the thermal continuum spectrum, radiative efficiency, and emissivity of the disk \cite{Shakura:1972te, Novikov:1973kta, Page:1974he}. Furthermore, upcoming observational experiments, including the EHT facility, will provide first-of-their-kind data on supermassive black holes like Sgr A*~\cite{EventHorizonTelescope:2022wkp}. We also highlight that such observational data help constrain the parameter space of NLED theories based on the black hole shadow and photon spheres.

Within this setup, the spacetime metric of the underlying black hole is specified by six parameters: mass of the black hole $m$, cosmological constant $\Lambda$, its charge $q$, magnetic charge $p$, and non-linear electrodynamics couplings $a$ and $b$. In order to fulfill the dominant energy condition, the coupling constants should be non-negative: $a \ge 0$, $b \ge 0$ \cite{Croney:2025cue}. In the field equations, the electrostatic part $F_{rt}$ is specified implicitly through a nontrivial cubic equation. In order to keep the non-linear corrections subdominant in the weak-field expansion, the following values for the couplings are typically used in numeric examples: $a \sim 10^{-4}{-}10^{-2}$ and $b \sim 10^{-3}{-}10^{-2}$, and small charges $q,p \sim 0.05{-}0.2$. In the purely magnetic case ($q=0$), the existence of the event horizon for an arbitrary value of the mass is assured if $p^2 < \frac{27}{2}a$, while in the case of $p^2 > \eta a$ with $\eta \approx 13.987$, an extremal solution appears \cite{Croney:2025cue}. Throughout the present analysis, the cosmological constant is restricted to $\Lambda \le 0$, i.e., to asymptotically anti-de Sitter (AdS) or flat geometries.

Throughout this work, we adopt geometrized units where $G=c=1$. To facilitate scale-independent calculations, we measure all physical quantities in terms of the black hole mass $m$. Accordingly, we map lengths, charges, and coupling parameters to dimensionless variables.
By virtue of these variables, all numerical computations are seamlessly performed with $m=1$ without any loss of generality.

Furthermore, this paper is organized as follows. Section (\ref {sec:2}) considers the basic dyonic black hole metric in the context of non-linear electrodynamics and the particular characteristics of its horizon. Section (\ref{sec:3}) contains the study of the geodesics of motion and the derivation of the effective potential for test particles. The physics of thin accretion disks along with their kinematics are considered in Section (\ref{sec:4}). On the other hand, Section (\ref{sec:5}) consists of a detailed numerical study centered around ISCO limits, radiative features, and comparisons with the EHT data on the Sgr A* shadow. Finally, we conclude in Section (\ref{sec:6}).

\section{The Dyonic Black Hole Metric in Non-Linear Electrodynamics}\label{sec:2}

In this section, we consider the static, spherically symmetric black hole, derived by Ref. \cite{Croney:2025cue}, and we evaluate the horizons of this black hole by considering $g^{rr}=0$.
In this paper, we focus on thin accretion disks around this black hole and related issues.

\begin{eqnarray}\label{Eq1}
d s^{2} = -f(r) d t^{2} + \frac{1}{f(r)} d r^{2} + r^{2} (d\theta^{2} + sin^{2}\theta d\varphi^{2}),
\end{eqnarray}

where

\begin{eqnarray}\label{Eq2}
f(r) = 1 - \frac{2m}{r} - \frac{\Lambda r^{2}}{3} + \frac{p^{2}}{r^{2}} - \frac{2 a p^{4}}{5 r^{6}} \nonumber\\ + \frac{1}{2r} \int^{r} d r^{'} \big[ \big(r^{'2}+ \frac{4 p^{2}}{r^{'2}}(2b-a) \big) F_{rt}^{2} - 3 q F_{rt} \big].
\end{eqnarray}

Notice that the $F_{rt}$ is the root of the cubic equation \cite{Croney:2025cue},

\begin{eqnarray}\label{Eq3}
4 a F_{rt}^{3} + \big( 1+ \frac{4 p^{2}}{r^{4}}(2b - a) \big) F_{rt} - \frac{q}{r^{2}} =0,
\end{eqnarray}

This is the most general form. In order to use it, we must numerically solve the above equation for $F_{rt}(r)$ at each radial step. Then, we should perform the integral in equation (\ref{Eq2}) to get $f(r)$. This approach is necessary to explore the full, rich parameter space.
\\

In a special case, we also consider the magnetic black hole case. This is a special, tractable case where the electric charge $q=0$.

\begin{eqnarray}\label{Eq4}
f(r) = 1 - \frac{2m}{r} - \frac{\Lambda r^{2}}{3} + \frac{p^{2}}{r^{2}} - \frac{2 a p^{4}}{5 r^{6}},
\end{eqnarray}

which is a simplification of equation (\ref{Eq2}) when $q=0$ and therefore $F_{rt}=0$ \cite{Croney:2025cue}. It's fully analytic and highlights the key new feature, the $-1/r^{6}$ term sourced by the magnetic charge $p$ and non-linear electrodynamics coupling $a$. We will study this case to show how non-linear electrodynamics modifies the strong-field region without the complexity of the cubic.

\begin{table}[htbp]\label{Table-01}
\centering
\caption{Parameter sets for black hole solutions.}
\begin{tabular}{lcccccc}
\toprule
\textbf{Label} & \(a\) & \(b\) & \(q\) & \(p\) & \(\Lambda\) & $Color$ \\
Schwarzschild         & 0      & 0      & 0    & 0    & 0 & Black \\
RN    & 0      & 0      & 0.6  & 0.1  & 0 & Blue \\
RN AdS & 0      & 0      & 0.3  & 0.1  & $-0.001$ & Purple \\
NLED 1                & 0.00037& 0.028  & 0.8  & 0.1  & 0 & Red\\
NLED 2                & 0.01   & 0.01   & 0.05 & 0.2  & 0 & Green \\
NLED AdS              & 0.003  & 0.028  & 0.15 & 0.1  & $-0.001$ & Orange \\
Magnetic 1                & 0.00037& 0.028  & 0  & 0.1  & 0 & Yellow\\
Magnetic 2                & 0.01   & 0.028   & 0 & 0.8  & 0 & Brown \\
Magnetic AdS              & 0.003  & 0.028  & 0 & 0.1  & $-0.001$ & Pink \\
\end{tabular}\label{Table-01}
\end{table}

To determine the different kinds of black hole parameters, we consider Table (\ref{Table-01}). The choices of the coupling constants $a, b$, electric charge $q$, magnetic charge $p$, and the cosmological constant $\Lambda$ used in our numerical analysis are shown in Table (\ref{Table-01}). These cases include general vacuum relativity (Schwarzschild), linear Maxwell field solutions (Reissner-Nordstr\"om in flat and anti-de Sitter spacetimes), and new cases involving dyonic or purely magnetic non-linear electrodynamics. To clearly label the plots showing the horizon structure, potential, and disk profile of each case, we use a unique color for each one.

\begin{figure}
\centering
\includegraphics[width=1\linewidth]{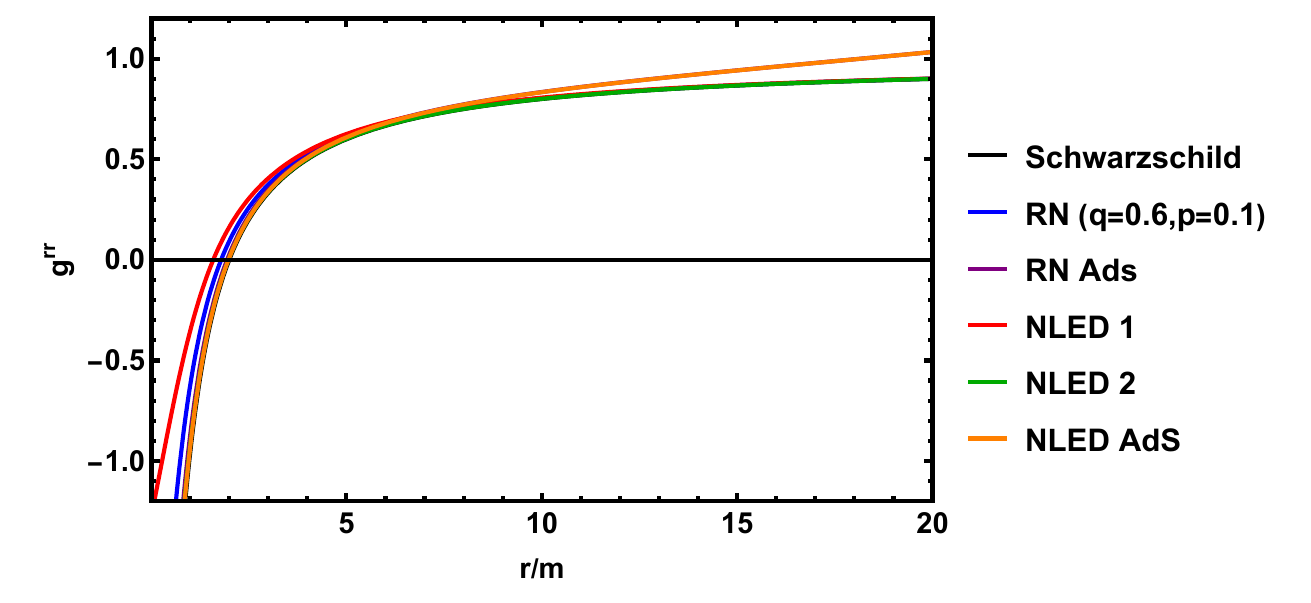}
\caption{The graphs show the profiles of $g^{rr}$ for the different kinds of black holes. The event horizon radius can be found by solving $g^{rr}=0$, and the number of crossings shows the number of horizons.}
\label{fig01}
\end{figure}

\begin{figure*}
\centering
\includegraphics[width=0.9\textwidth]{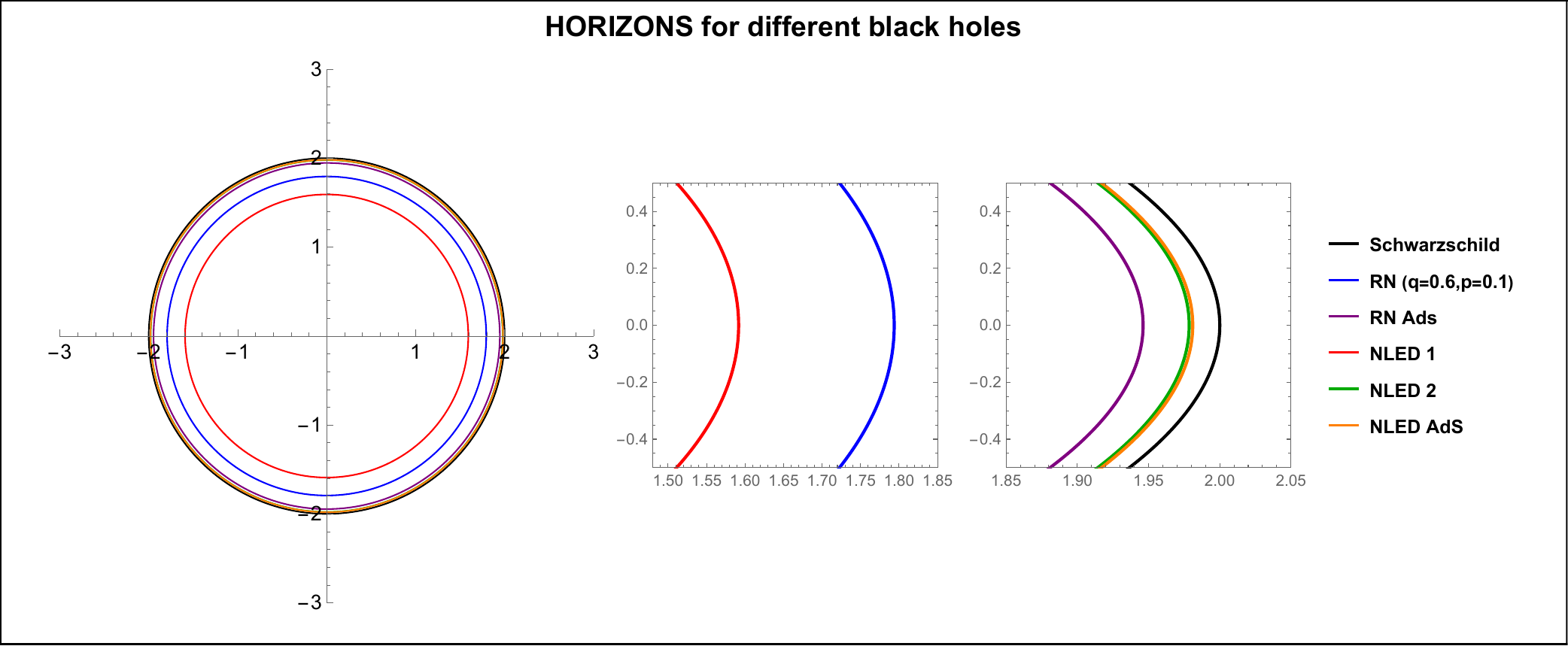}
\caption{The graph shows the event horizons for the different kinds of black holes: the Schwarzschild, Reissner-Nordstr\"om, and dyonic black holes.}
\label{fig02}
\end{figure*}

Structural properties and position of event horizons for various black holes are analyzed in Figs.~(\ref{fig01}) and (\ref{fig02}). Fig. (\ref{fig01}) shows the radial dependence of the metric function components $g^{rr}$ across varying models to find the radius and multiplicity of horizons, with roots $g^{rr}=0$ corresponding to physically relevant horizon boundaries. In order to make a better geometrical illustration of the results, Fig.~(\ref{fig02}) depicts concentric event horizons for the standard Schwarzschild, Reissner-Nordstr\"om, and new NLED dyonic solutions. As the lines show, NLED effects and charge coupling destroy Maxwellian linear duality and cause the outer horizon to contract and shift inward compared with standard general relativity counterparts.

\section{Geodesic Motion and Fundamental Orbital Dynamics}\label{sec:3}

In this stage, we derive the equations of motion and the effective potential to evaluate the system's dynamics. Thus, we can consider the Lagrangian $\mathcal{L}$ for a point particle around the dyonic black hole in non-linear electrodynamics, and this metric is given by Eq. (\ref{Eq1}). The Lagrangian is
\begin{eqnarray}
\mathcal{L}=\frac{1}{2}g_{\mu\nu}\frac{dx^{\mu}}{ds}\frac{dx^{\nu}}{ds}=\frac{1}{2}\varepsilon.
\end{eqnarray}

Notice that $\varepsilon =1$ shows a massive particle, and $\varepsilon =0$ can be considered for the photon. One can consider this issue in the equatorial plane; $E$ is conserved energy, and $L$, angular momentum, can be obtained as follows
\begin{align}
E &= - g_{tt}\frac{dt}{ds} = f(r)\frac{dt}{ds}, \label{eq:E}
\end{align}

\begin{eqnarray}
L=g_{\varphi\varphi}\frac{d\varphi}{ds}=r^{2}\frac{d\varphi}{ds}.
\end{eqnarray}

Thus, the geodesic equations of a massive particle can be obtained 

\begin{eqnarray}\label{E-M1}
\big(\frac{dr}{ds}\big)^{2}=E^{2}- f(r) \big(1+\frac{L^{2}}{r^{2}}\big), \nonumber\\
\end{eqnarray}

\begin{eqnarray}\label{E-M2}
\big(\frac{dr}{d\varphi}\big)^{2}=\frac{r^{4}}{L^{2}}\big\lbrace E^{2}- f(r) \big(1+\frac{L^{2}}{r^{2}}\big)\big\rbrace, \nonumber\\
\end{eqnarray}

\begin{eqnarray}\label{E-M3}
\big(\frac{dr}{dt}\big)^{2}=\frac{1}{E^{2}} f(r)^{2}\bigg\lbrace E^{2}- f(r) \big(1+\frac{L^{2}}{r^{2}}\big)\bigg\rbrace. \nonumber\\
\end{eqnarray}

Using Eqs. (\ref{E-M1}-\ref{E-M3}), we can fully describe the system dynamics. Moreover, the effective potential can be obtained from Eq. (\ref{E-M1}) as follows

\begin{eqnarray}\label{V_eff1}
V_{eff}= f(r) \big(1+\frac{L^{2}}{r^{2}}\big).
\end{eqnarray}

To facilitate the analysis and exploit the equations' scaling properties, we introduce dimensionless quantities by scaling with appropriate powers of the black hole mass $m$. We define
\begin{equation}
\begin{array}{rcl}
\bar{r} = \dfrac{r}{m}, &\quad \bar{\Lambda} = \Lambda m^{2}, &\quad \bar{L} = \dfrac{L}{m},\\[8pt]
\bar{a} = \dfrac{a}{m^{2}}, &\quad \bar{b} = \dfrac{b}{m^{2}}, &\quad \bar{q} = \dfrac{q}{m},\quad \bar{p} = \dfrac{p}{m}.
\end{array}
\label{eq:dimensionless}
\end{equation}

In these variables, all equations of motion become independent of the overall mass scale $m$. Consequently, we can assure that our computations are reliable without loss of generality, and the results for any physical black hole are then recovered by multiplying the dimensionless quantities by the appropriate power of $m$.

In all numerical calculations, we set the black hole mass to unity, $m=1$, so that the dimensionless quantities $\bar{r}, \bar{t}, \bar{L}, \bar{a}, \bar{b}, \bar{q}, \bar{p},$ and $\bar{\Lambda}$ completely determine the dynamics. Physical observables for an astrophysical black hole of physical mass $m_{\text{phys}}$ at distance $D_{\text{phys}}$ are obtained by multiplying the dimensionless results by the appropriate powers of $m_{\text{phys}}$ (since $G=c=1$ implies mass, length, and time have the same dimension).

Now, by considering Eq. (\ref{eq:dimensionless}) we have

\begin{eqnarray}\label{V_eff}
\bar{V}_{eff}= \bar{f}(\bar{r}) \big(1+\frac{\bar{L}^{2}}{\bar{r}^{2}}\big),
\end{eqnarray}

\begin{figure}
\centering
\includegraphics[width=1\linewidth]{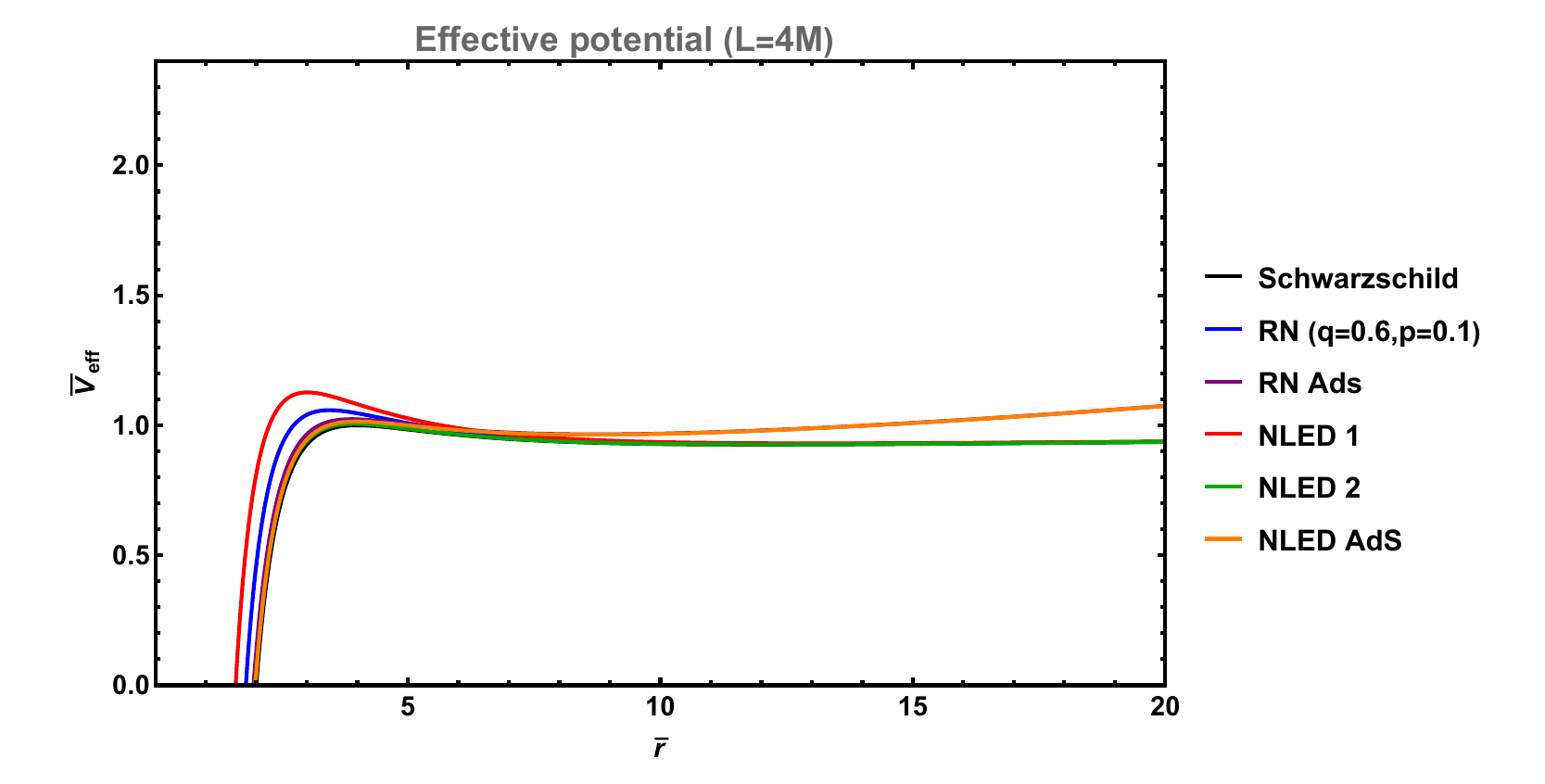}
\caption{The graph shows the effective potentials for the Schwarzschild, Reissner-Nordstr\"om, and dyonic black holes.}
\label{fig03}
\end{figure}

The nature of the effective potential for the motion of test particles in the mentioned spacetimes is depicted in Fig.~(\ref{fig03}). In this figure, the radial distribution of the effective potential is shown for the Schwarzschild, Reissner-Nordstr\"om, and non-linear electrodynamics dyonic black holes. The charge and the nonlinearity of the electromagnetic field affect the effective potential barriers and their peak locations, which in turn changes the stability threshold and the ISCO radius in each black hole scenario.

\subsection{The Magnetic Black Hole}

In this special case, we consider the magnetic black hole case (\ref{Eq4}), the dimensionless metric function becomes

\begin{eqnarray}\label{Eq14}
\bar{f}_{\text{Mag}}(\bar{r}) = 1 - \frac{2}{\bar{r}} - \frac{\bar{\Lambda}}{3}\,\bar{r}^{2} + \frac{\bar{p}^{2}}{\bar{r}^{2}} - \frac{2\bar{a}\bar{p}^{4}}{5\bar{r}^{6}}.
\end{eqnarray}

\begin{figure}
\centering
\includegraphics[width=1\linewidth]{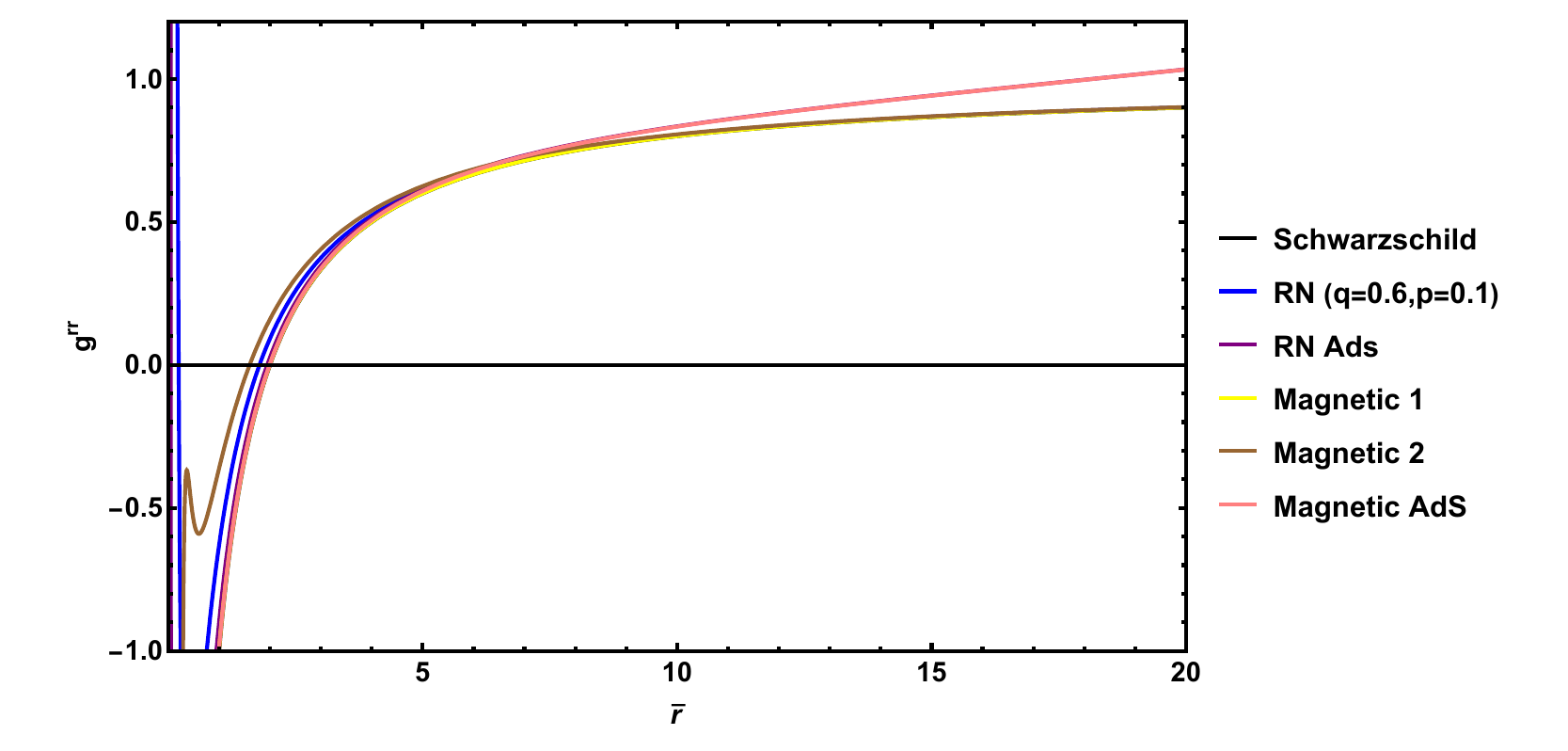}
\caption{The graphs show the profiles of $g^{rr}$ for the Magnetic black hole.}
\label{fig04}
\end{figure}

\begin{figure*}
\centering
\includegraphics[width=0.8\textwidth]{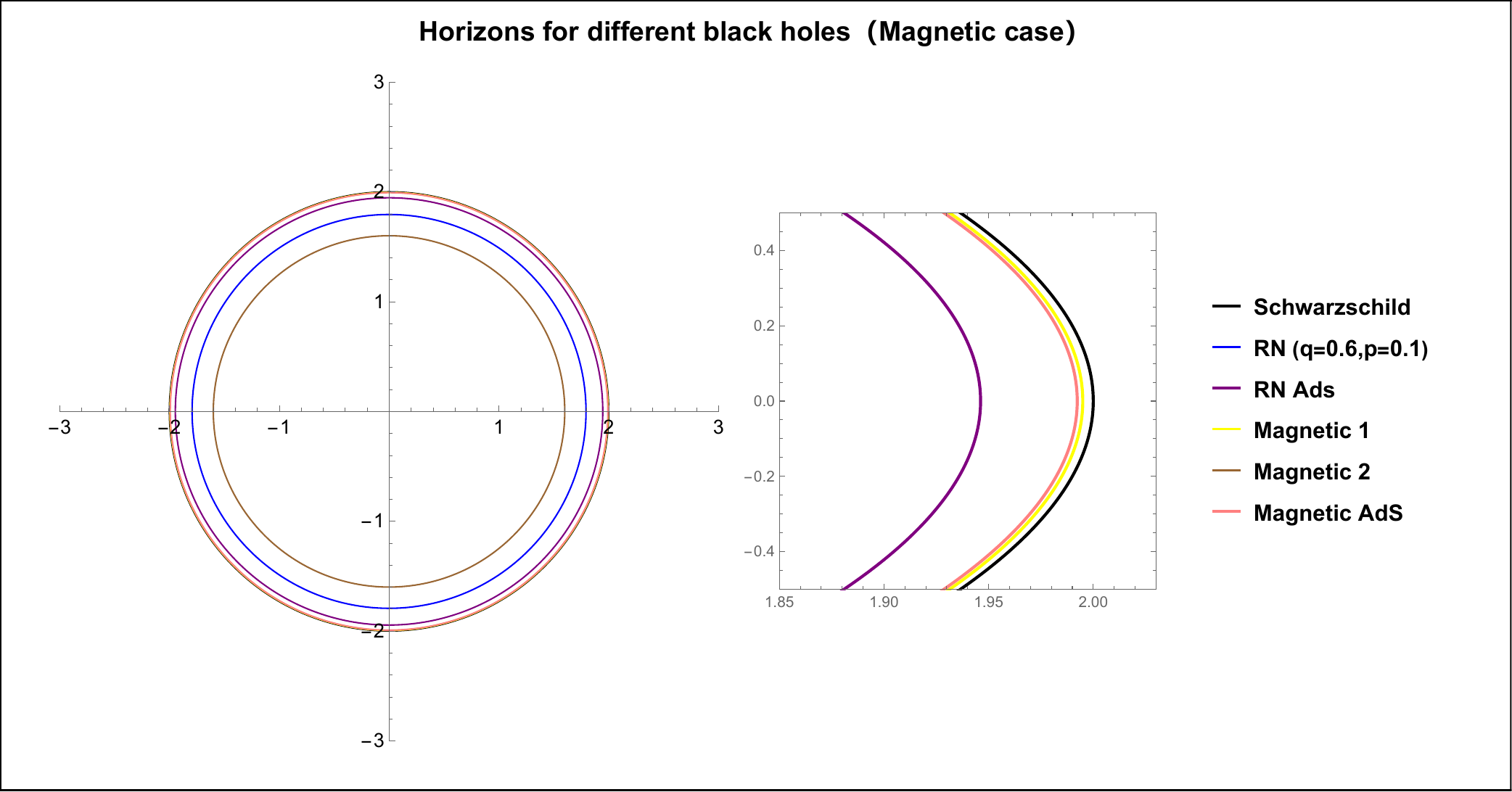}
\caption{The graph shows the event horizons for different kinds of black holes: the Schwarzschild, Reissner-Nordstr\"om, and magnetic black holes.}
\label{fig05}
\end{figure*}

Properties of the structure and geometrical positions of the horizons of the black hole solution, which is purely magnetic in nature, are provided in the Figs. (\ref{fig04}) and (\ref{fig05}). More specifically, in Fig.~(\ref{fig04}), the dependence of the metric function element $g^{rr}$ provided in Eq.~\eqref{Eq14} on the coordinate $\bar{r}$ is shown; here we note the impact of changes in the values of the parameters on the existence of multiple roots of $g^{rr}=0$. In order to give some geometric perspective in comparison, Fig.~(\ref{fig05}) shows the concentric horizons along with the usual Schwarzschild and Reissner-Nordstr\"om horizons. The effect of the non-linear correction term in electrodynamics, i.e., $\bar{r}^{-6}$ in Eq.~\eqref{Eq14}, becomes clear.

For a massive particle ($\varepsilon=1$) in the equatorial plane, the geodesic equations~\eqref{E-M1}--\eqref{E-M3} become

\begin{align}
\left(\frac{d\bar{r}}{d\bar{s}}\right)^{2}_{\text{Mag}} &= \bar{E}^{2} - \bar{f}_{\text{Mag}}(\bar{r})\left(1+\frac{\bar{L}^{2}}{\bar{r}^{2}}\right), \label{drds_Mag}\\
\left(\frac{d\bar{r}}{d\phi}\right)^{2}_{\text{Mag}} &= \frac{\bar{r}^{4}}{\bar{L}^{2}}\left[ \bar{E}^{2} - \bar{f}_{\text{Mag}}(\bar{r})\left(1+\frac{\bar{L}^{2}}{\bar{r}^{2}}\right) \right], \label{drdphi_Mag}\\
\left(\frac{d\bar{r}}{d\bar{t}}\right)^{2}_{\text{Mag}} &= \frac{\bar{f}_{\text{Mag}}(\bar{r})^{2}}{\bar{E}^{2}}\left[ \bar{E}^{2} - \bar{f}_{\text{Mag}}(\bar{r})\left(1+\frac{\bar{L}^{2}}{\bar{r}^{2}}\right) \right]. \label{drdt_Mag}
\end{align}

The effective potential is

\begin{eqnarray}\label{Veff_Mag}
\bar{V}_{\text{eff,Mag}}(\bar{r}) = \bar{f}_{\text{Mag}}(\bar{r})\left(1+\frac{\bar{L}^{2}}{\bar{r}^{2}}\right).
\end{eqnarray}

\begin{figure}
\centering
\includegraphics[width=0.5\textwidth]{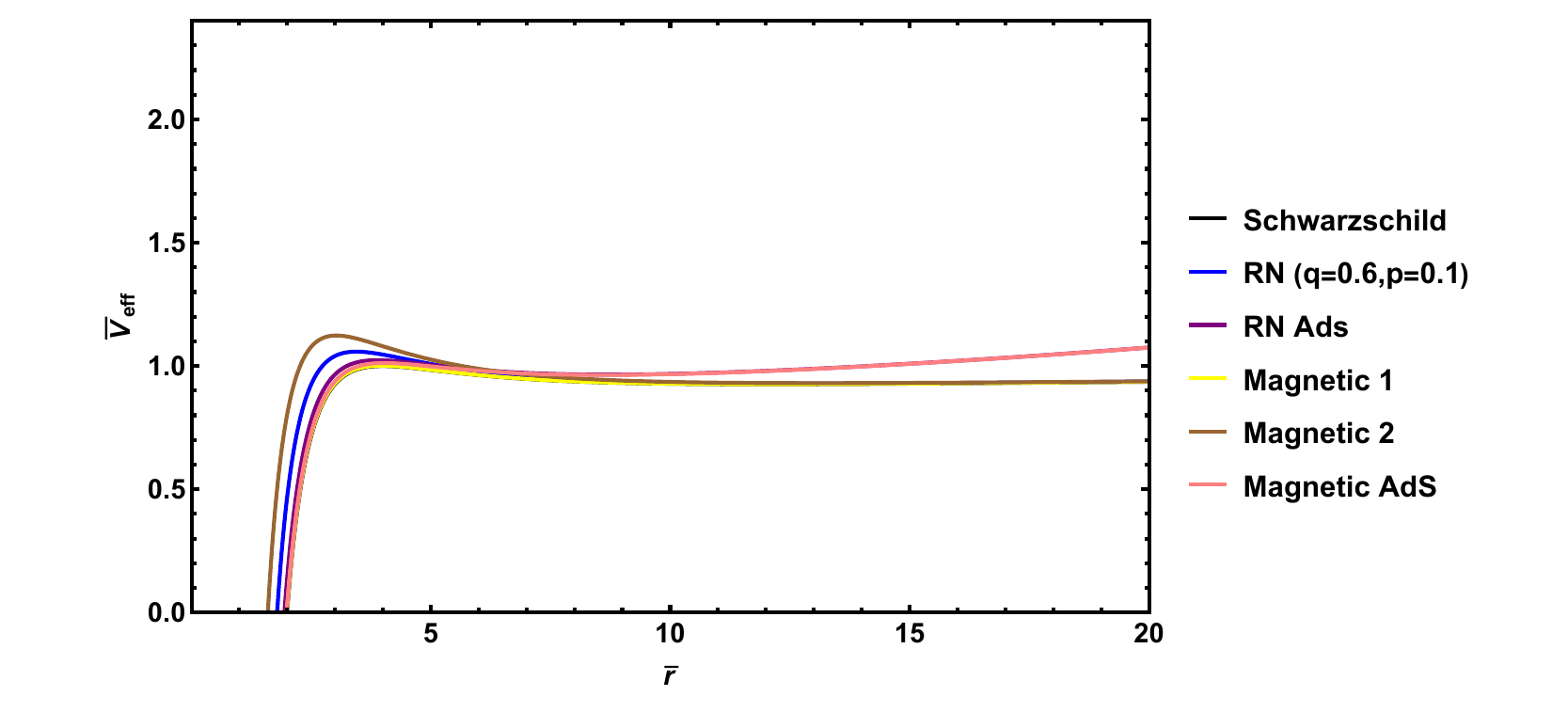}
\caption{The graph shows the effective potentials for the Schwarzschild, Reissner-Nordstr\"om, and Magnetic black holes.}
\label{fig06}
\end{figure}

For the radial distribution of the effective potential $\bar{V}_{\text{eff, Mag}}(\bar{r})$ (given in Eq.~\eqref{Veff_Mag}), we plot in Fig.~(\ref{fig06}) the results for test particles that move on the equatorial plane. In this figure, the effect of the magnetic charge correction is studied in comparison with the potentials of the Schwarzschild and Reissner-Nordstr\"om black holes. 

\section{Physics of the Thin Accretion Disk}\label{sec:4}

In the subsequent sections, we will discuss all the parameters of the accretion disk of the dyonic black hole in non-linear electrodynamics. \\
It is worth noting that the specific energy $\bar{E}$, specific angular momentum $\bar{L}$, angular velocity $\bar{\Omega}$, and the flux of the radiant energy $F$, in the disc of particles which rotate in a circular orbit, are known in all cases. \\ It should be mentioned that the physical parameters of the accretion disk result from certain structural equations, which are related to conservation of mass, energy, and angular momentum. As can be seen, the kinematic parameters depend on the orbital radius, and they can be found using general relations, which have been proposed by \cite{Novikov:1973kta,Thorne:1974ve,Harko:2009xf,Page:1974he}. Note that the calculation of all parameters is carried out using the dimensionless quantities which are defined in Eq. (\ref{eq:dimensionless}).

\begin{align}
\bar{\Omega}  &= \sqrt{-\frac{g_{tt,r}}{g_{\phi\phi,r}}} = \sqrt{\frac{\bar{f'}(\bar{r})}{2 \bar{r}}}, \label{eq:Omega}\\
\bar{E} &= -\frac{g_{tt}}{\sqrt{-g_{tt} - g_{\phi\phi}\Omega^2}} = \frac{\bar{f}(\bar{r})}{\sqrt{\bar{f}(\bar{r}) - \frac{\bar{r}}{2}\bar{f'}(\bar{r})}}, \label{eq:E_circ}\\
\bar{L} &= \frac{g_{\phi\phi}\Omega}{\sqrt{-g_{tt} - g_{\phi\phi}\Omega^2}} = \sqrt{\frac{\bar{f'}(\bar{r})\, \bar{r}^3}{2\bar{f}(\bar{r}) - \bar{r} \bar{f'}(\bar{r})}}. \label{eq:L_circ}
\end{align}

We can consider the conditions $\bar{V}_{eff}=0$ and $\frac{d\bar{V}_{eff}}{d\bar{r}}=0$ since we need to apply the conditions of the effective potential in order to make the particles travel in circular paths. So, from equation (\ref{V_eff}),

\begin{equation}
\frac{d^2\bar{V}_{\mathrm{eff}}}{d\bar{r}^{2}}\bigg|_{\bar{r}=\bar{r}_{\mathrm{ISCO}}} = 0 \quad\Longrightarrow\quad 
\bar{r}\,\bar{f}\,\bar{f}'' + 3\bar{f}\,\bar{f}' - 2\bar{r}\,(\bar{f}')^{2} = 0, \label{eq:ISCO_condition}
\end{equation}

Notice that innermost circular orbits occur at the local minimum of the effective potential; the dimensionless radius of the innermost stable circular geodesic orbit is obtained from $\frac{d^{2}\bar{V}_{eff}}{d\bar{r}^{2}}=0$.

Thus, the flux of the radiant energy over the disk can be obtained from the following relation \cite{Page:1974he}.

\begin{align}\label{Eq19}
F (r)=\frac{-\dot{M}_{0}}{4\pi\sqrt{-g}}\frac{\bar{\Omega}_{,r}}{(\bar{E}-\bar{\Omega} \bar{L})^{2}}\int_{r_{ISCO}}^{r}(\bar{E}-\bar{\Omega} \bar{L})\bar{L}_{,r}dr,
\end{align}

which follows the Stefan–Boltzmann law when the disk is in thermal equilibrium.
Here, the $\dot{M}_{0}$ is the mass accretion rate. 
It should be mentioned that, using the conservation laws, the flux of radiant energy over the disk spanning between the ISCO and a certain radial distance should be achieved \cite{Page:1974he,Gyulchev:2019tvk}.
\newpage

\begin{figure}
\centering
\includegraphics[width=0.5\textwidth]{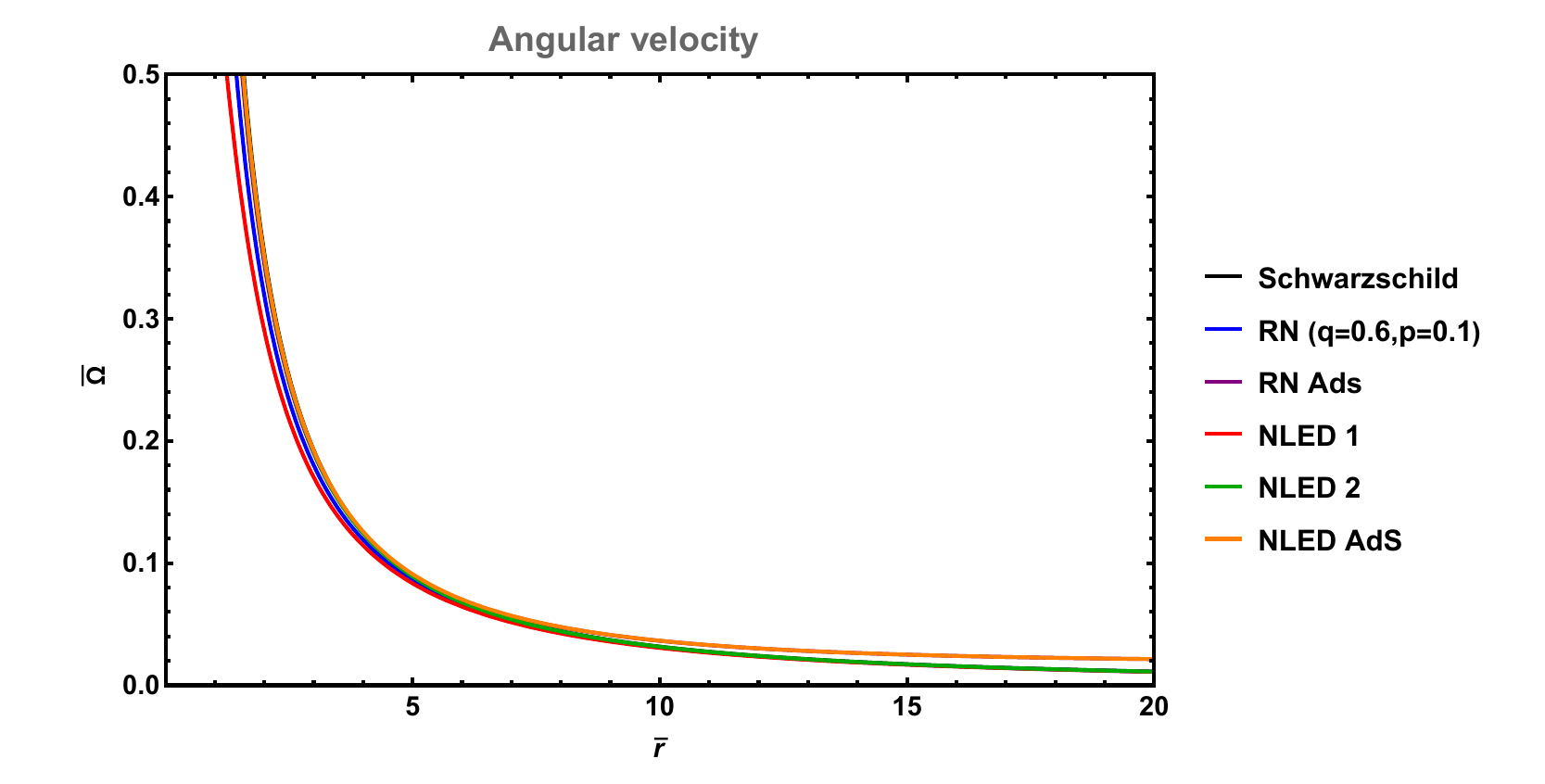}
\caption{The graphs show the angular velocity $\bar{\Omega}$ as a function of $\bar{r}$ for the particles moving around the black hole and forming the thin accretion disk.}
\label{fig7}
\end{figure}

\begin{figure}
\centering
\includegraphics[width=0.5\textwidth]{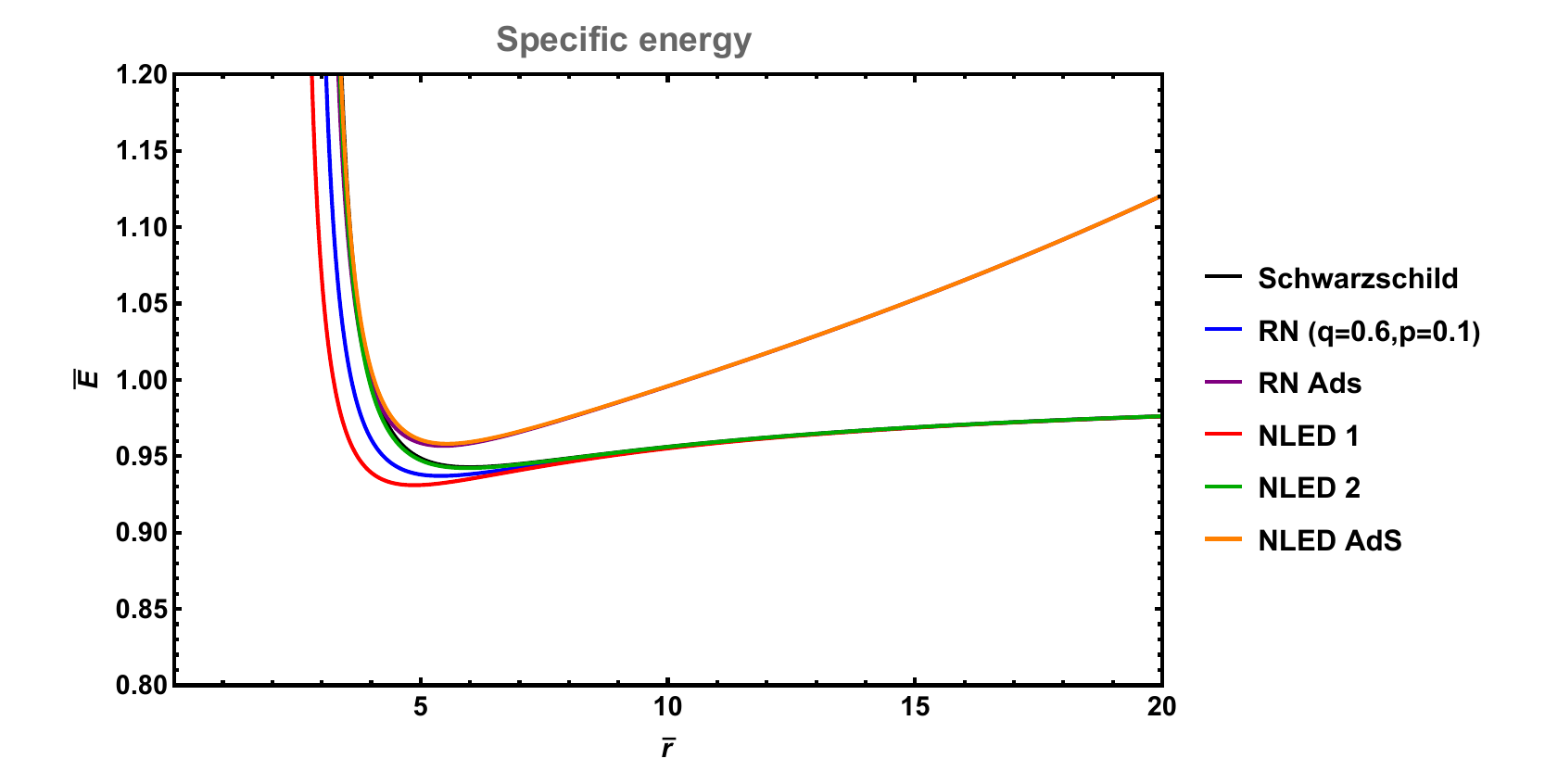}
\caption{The graphs show the specific energy $\bar{E}$ as a function of $\bar{r}$ for the particles moving around the black hole and forming the thin accretion disk.}
\label{fig8}
\end{figure}

\begin{figure}
\centering
\includegraphics[width=0.5\textwidth]{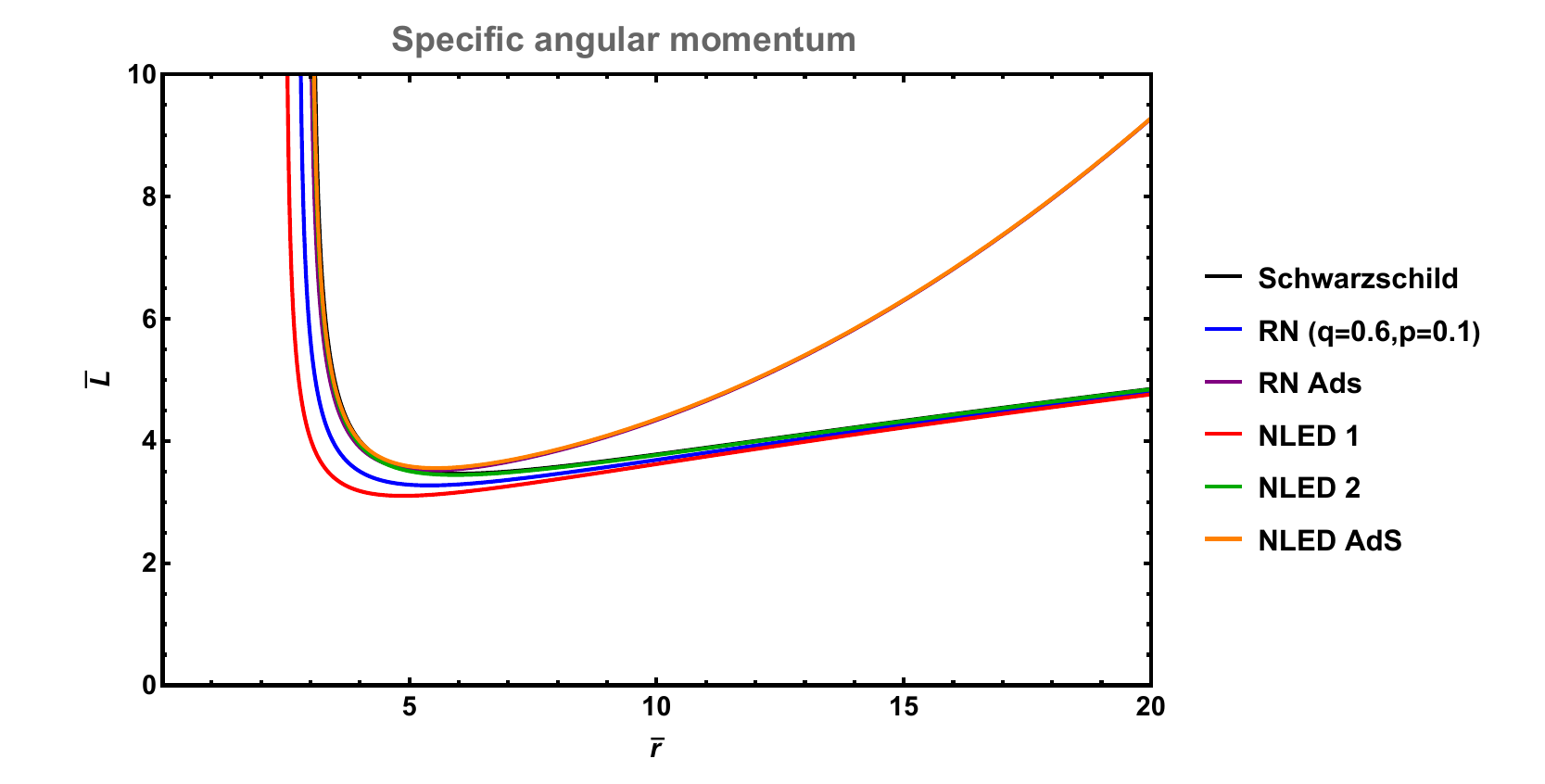}
\caption{The graphs show the specific angular momentum $\bar{L}$ versus $\bar{r}$ for the particles moving around the black hole and making the thin accretion disk.}
\label{fig9}
\end{figure}

\begin{figure}
\centering
\includegraphics[width=0.5\textwidth]{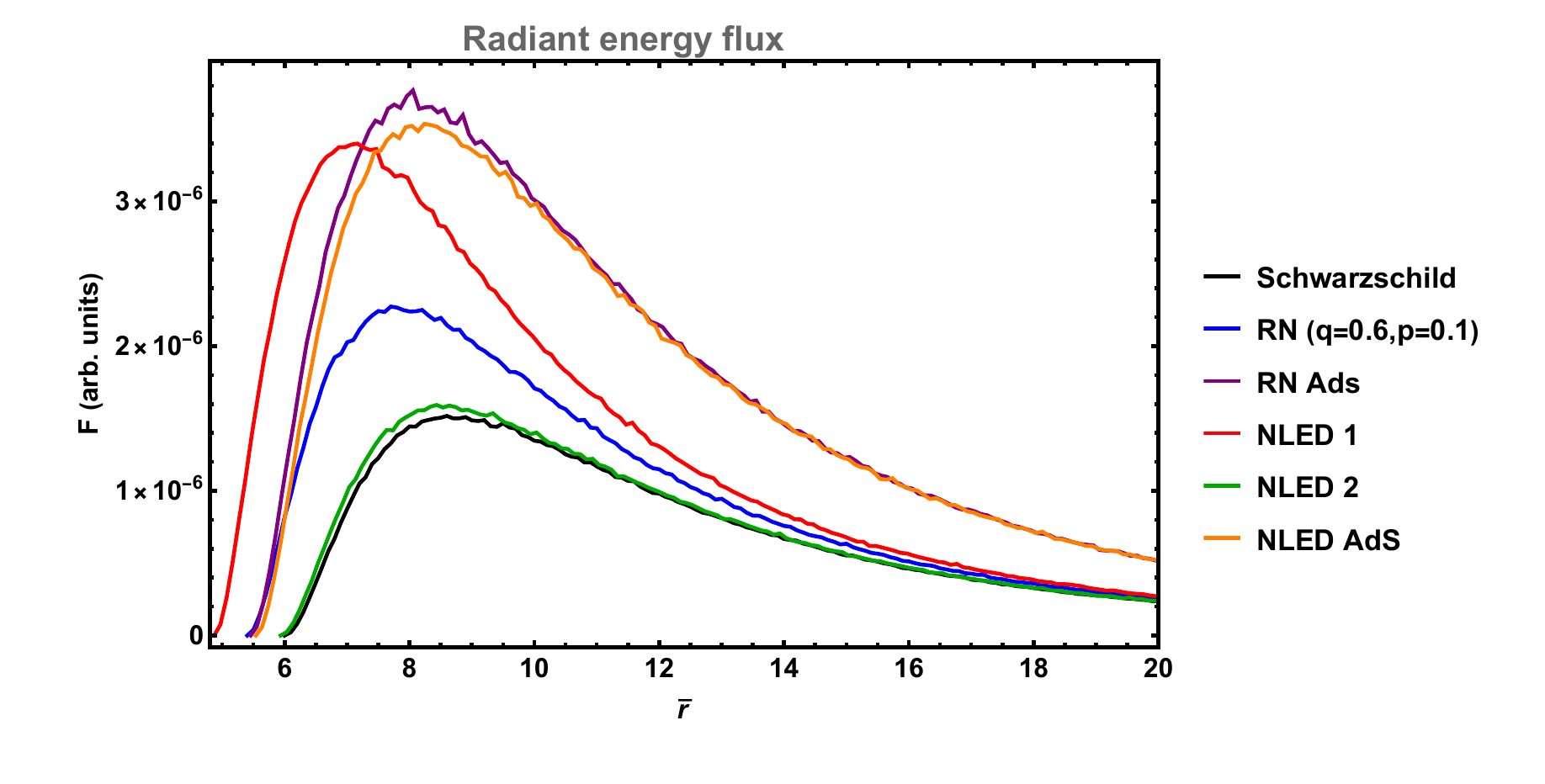}
\caption{The graphs show the flux of the radiant energy $F(r)$ for the particles moving around the black hole and making the thin accretion disk.}
\label{fig10}
\end{figure}

In Fig (\ref{fig7}), we have presented the angular velocity $\Omega$ with respect of $\bar{r}$. In fact, we have illustrated the angular velocity for different black hole models with different values. As shown in the figure, the angular velocity increases as particles move toward the black holes, reaching high values near the event horizons. It is obvious that the angular velocity has a similar pattern for all of the cases of black holes.

In Fig (\ref{fig8}), the specific energy $\bar{E}$ of the particles of the thin accretion disk at $\bar{r}$ is demonstrated for different kinds of black holes with different values of parameters. From an overall perspective, the most striking feature of Fig(\ref{fig8}) is that the specific energy exhibits three phases. The first is far from the black hole, the second is near the black hole, and the third is close to the event horizon. The particles' specific energy decreases as they approach the black hole, reaching a minimum. Then, as they approach the event horizon, they reach their highest value and rise sharply. It should be pointed out that the behavior of the specific energy for the Reissner-Nordstr\"om AdS (RN AdS) and the Non-Linear Electrodynamics AdS (NLED AdS) is the same.

In Fig (\ref{fig9}), the graphs show the specific angular momentum $\bar{L}$ as a function of $\bar{r}$ for different cases of black holes. We can immediately see that the specific angular momentum is very similar to the specific energy for all cases.

In Fig (\ref{fig10}), the graphs indicate the flux of the radiant energy $F(r)$ for the different black holes. According to the graphs, the radiant energy flux grows gradually as particles move toward the black hole. Furthermore, it reaches a maximum value and then falls suddenly near the event horizon. Notice that Schwarzschild has the lowest values, and the Reissner-Nordstr\"om AdS (RN AdS) has the highest values. The other cases fall between them.

It should be mentioned that these kinematic quantities are intimately connected through the location of the ISCO. According to the stability analysis, the local minima of the specific energy $\bar{E}$ and the specific angular momentum $\bar{L}$ define the exact radius of the innermost stable circular orbit. At this inner boundary, the radiant energy flux vanishes due to the zero-torque condition at the ISCO. The flux then reaches its peak at a slightly larger radius before falling off asymptotically at large distances. In contrast, the angular velocity $\bar{\Omega}$ does not exhibit a minimum and instead increases monotonically as the particles approach the horizon. These features are consistent with the standard picture of a relativistic thin accretion disk and provide a useful consistency check for our numerical results.

Therefore, we have obtained all parameters of the accretion disk of a dyonic black hole in non-linear electrodynamics. Moreover, we analyzed how these parameters change with different values of the theory's components.

\subsection{The Magnetic Black Hole}

For circular equatorial orbits we use the general formulas~\eqref{eq:Omega}--\eqref{eq:L_circ}.  First compute the derivative of $\bar{f}_{\text{Mag}}$:

\begin{align}
\bar{f}_{\text{Mag}}'(\bar{r}) &=
\frac{2}{\bar{r}^{2}} - \frac{2\bar{\Lambda}}{3}\,\bar{r} - \frac{2\bar{p}^{2}}{\bar{r}^{3}} + \frac{12\bar{a}\bar{p}^{4}}{5\bar{r}^{7}}. \label{fprime_Mag}
\end{align}

Then the angular velocity $\bar{\Omega}_{\text{Mag}}$, specific energy $\bar{E}_{\text{Mag}}$ and specific angular momentum $\bar{L}_{\text{Mag}}$ are

\begin{widetext}
\begin{align}
\bar{\Omega}_{\text{Mag}}(\bar{r}) &= \sqrt{\frac{\bar{f}_{\text{Mag}}'(\bar{r})}{2\bar{r}}}
= \sqrt{ \frac{1}{\bar{r}^{3}} - \frac{\bar{\Lambda}}{3} - \frac{\bar{p}^{2}}{\bar{r}^{4}} + \frac{6\bar{a}\bar{p}^{4}}{5\bar{r}^{8}} }, \label{Omega_Mag}\\[4pt]
\bar{E}_{\text{Mag}}(\bar{r}) &=
\frac{\bar{f}_{\text{Mag}}(\bar{r})}
{\sqrt{ \bar{f}_{\text{Mag}}(\bar{r}) - \frac{\bar{r}}{2}\bar{f}_{\text{Mag}}'(\bar{r}) }}
= \frac{\bar{f}_{\text{Mag}}(\bar{r})}
{\sqrt{ 1 - \dfrac{3}{\bar{r}} + \dfrac{2\bar{p}^{2}}{\bar{r}^{2}} - \dfrac{8\bar{a}\bar{p}^{4}}{5\bar{r}^{6}} }}, \label{E_Mag}\\[4pt]
\bar{L}_{\text{Mag}}(\bar{r}) &=
\sqrt{ \frac{ \bar{f}_{\text{Mag}}'(\bar{r})\,\bar{r}^{3} }
{ 2\bar{f}_{\text{Mag}}(\bar{r}) - \bar{r}\bar{f}_{\text{Mag}}'(\bar{r}) } }
= \sqrt{ \frac{ \bar{f}_{\text{Mag}}'(\bar{r})\,\bar{r}^{3} }
{ 2\left(1 - \dfrac{3}{\bar{r}} + \dfrac{2\bar{p}^{2}}{\bar{r}^{2}} - \dfrac{8\bar{a}\bar{p}^{4}}{5\bar{r}^{6}}\right) } }. \label{L_Mag}
\end{align}
\end{widetext}

\begin{figure}
\centering
\includegraphics[width=0.5\textwidth]{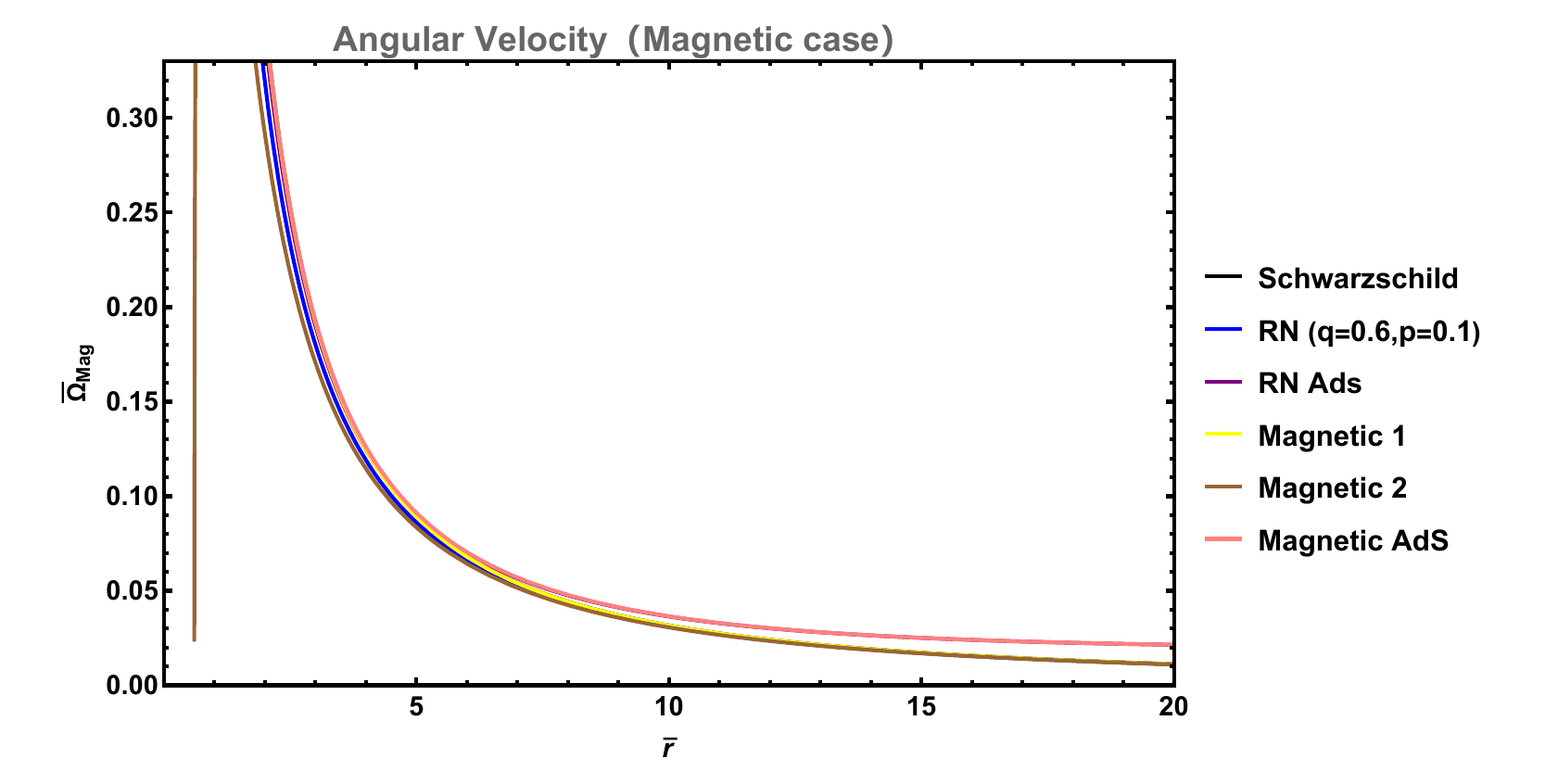}
\caption{The graphs show the angular velocity $\bar{\Omega}_{Mag}$ as a function of $\bar{r}$ for the particles moving around the Magnetic black hole and forming the thin accretion disk.}
\label{fig4Mg}
\end{figure}

\begin{figure}
\centering
\includegraphics[width=0.5\textwidth]{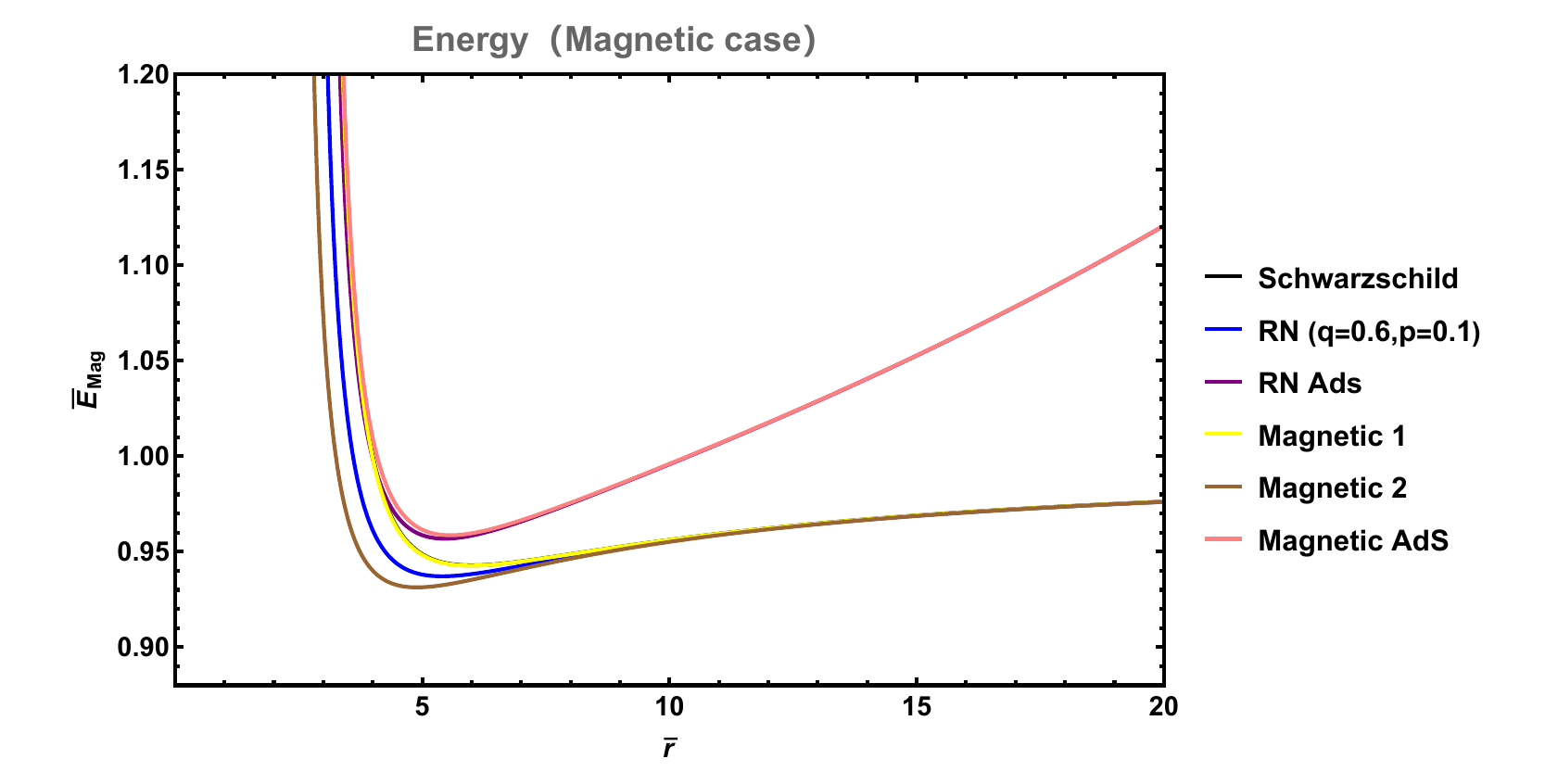}
\caption{The graphs show the specific energy $\bar{E}_{Mag}$ as a function of $\bar{r}$ for the particles moving around the Magnetic black hole and forming the thin accretion disk.}
\label{fig5Mg}
\end{figure}

\begin{figure}
\centering
\includegraphics[width=0.5\textwidth]{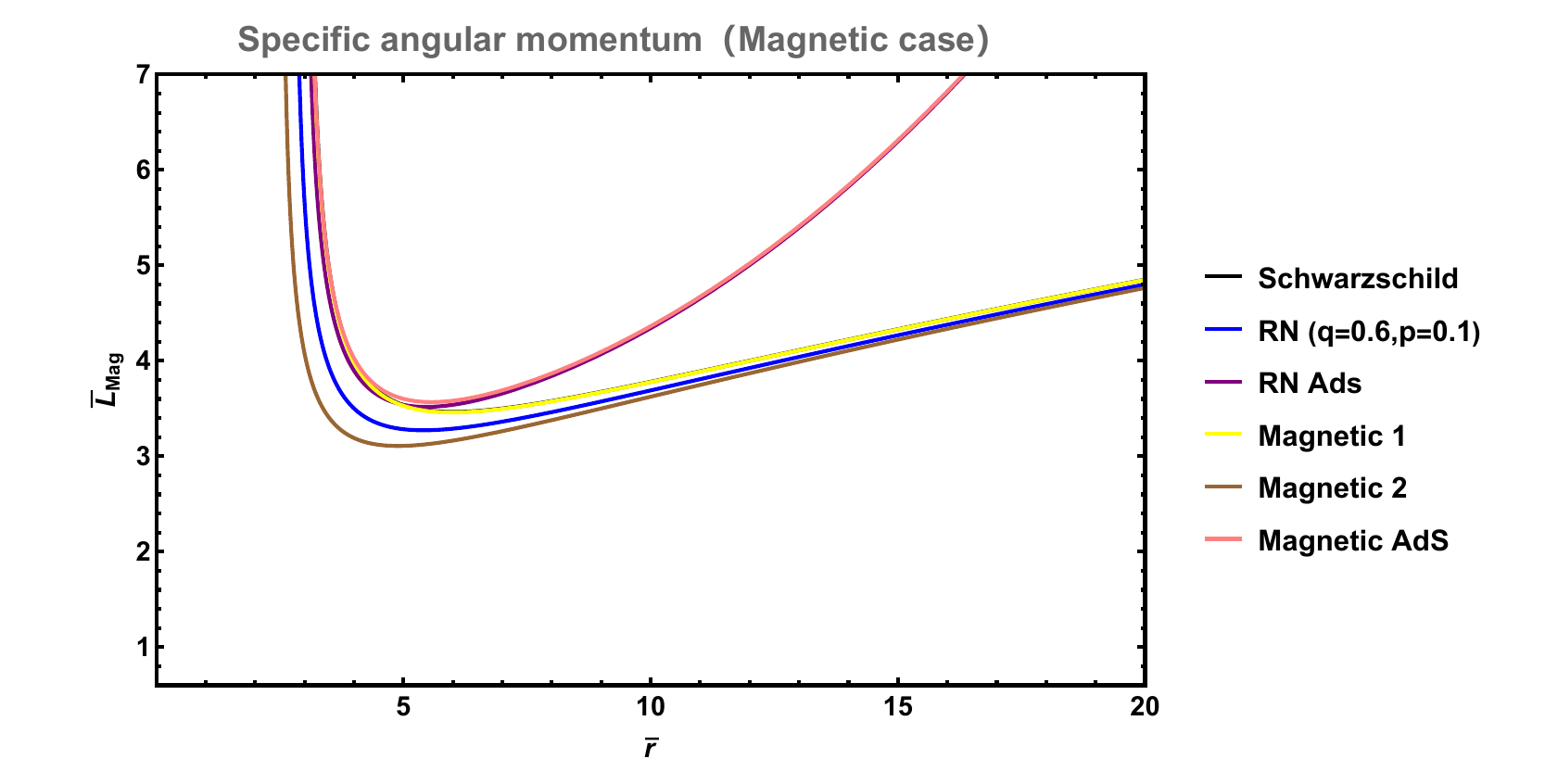}
\caption{The graphs show the specific angular momentum $\bar{L}_{Mag}$ as a function of $\bar{r}$ for the particles moving around the Magnetic black hole and forming the thin accretion disk.}
\label{fig6Mg}
\end{figure}

Basic characteristics of geodesic motion of the test particle forming the thin disk of accretion around the magnetic black hole are shown in Figs. (\ref{fig4Mg}), (\ref{fig5Mg}), and (\ref{fig6Mg}). In particular, the dependence of the angular velocity of the test particle on the dimensionless radial coordinate for different values of the non-linear magnetic charge is shown in Fig. (\ref{fig4Mg}). The specific energy of the test particle in the orbits of the magnetic black hole, $\bar{E}_{\text{Mag}}$, and the specific angular momentum, $\bar{L}_{\text{Mag}}$, are presented in Figs. (\ref{fig5Mg}) and (\ref{fig6Mg}), respectively. These dependences define the inner boundary of the accretion disk and the amount of binding energy used to emit radiation.

The denominator in $\bar{E}_{\text{Mag}}$ and $\bar{L}_{\text{Mag}}$ is positive for stable orbits outside the ISCO.

The energy flux from the thin accretion disk is given by~\eqref{Eq19}.  For the magnetic black hole, it becomes

\begin{widetext}
\begin{eqnarray}\label{flux_Mag}
F_{\text{Mag}}(\bar{r}) = \frac{-\dot{M}_{0}}{4\pi\sqrt{-g}}\;
\frac{\bar{\Omega}_{\text{Mag}}'(\bar{r})}
{\bigl(\bar{E}_{\text{Mag}}(\bar{r})-\bar{\Omega}_{\text{Mag}}(\bar{r})\bar{L}_{\text{Mag}}(\bar{r})\bigr)^{2}}
\int_{\bar{r}_{\text{ISCO}}}^{\bar{r}}
\bigl(\bar{E}_{\text{Mag}}(\bar{r})-\bar{\Omega}_{\text{Mag}}(\bar{r})\bar{L}_{\text{Mag}}(\bar{r})\bigr)\,
\bar{L}_{\text{Mag}}(\bar{r})\,d\bar{r},
\end{eqnarray}
\end{widetext}

with $\sqrt{-g}= \bar{r}^{2}\sin\theta$ for the metric~\eqref{Eq1}. The integral is evaluated numerically once $\bar{r}_{\text{ISCO}}$ is known. 

\begin{figure}
\centering
\includegraphics[width=0.5\textwidth]{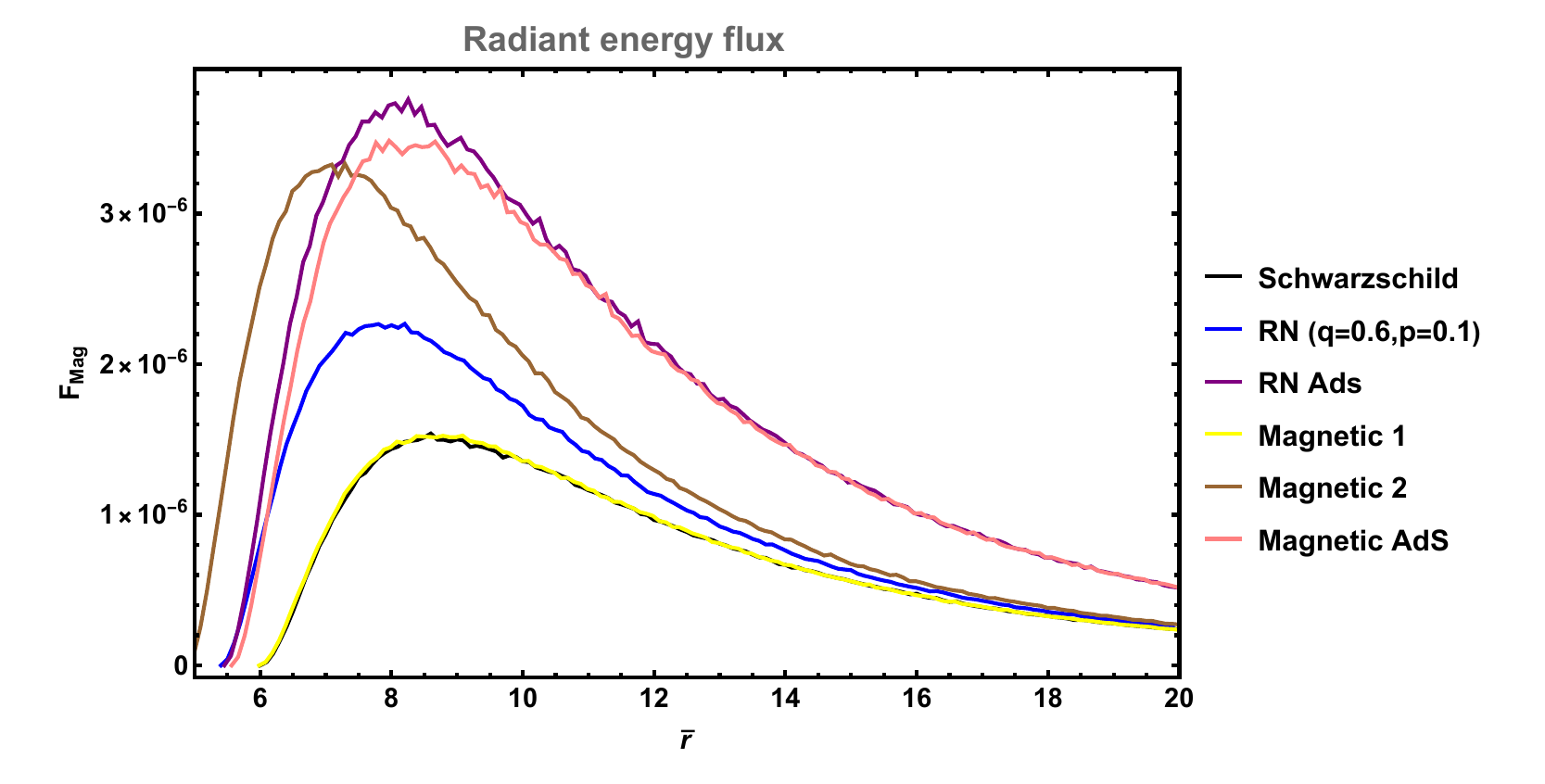}
\caption{The graphs show the flux of the radiant energy $F_{Mag}(r)$ for the particles moving around the magnetic black hole and forming the thin accretion disk.}
\label{fig7Mg}
\end{figure}

The radiation behavior of the thin accretion disk around the magnetic black hole is shown in Fig. (\ref{fig7Mg}). The profiles depict the variation of the radiative flux $F_{\text{Mag}}(\bar{r})$ obtained numerically via Eq. (\ref{flux_Mag}). The radiative flux becomes zero at the inner radius of the disk, i.e., the ISCO radius $\bar{r}_{\text{ISCO}}$, rises to the peak value very sharply, and then falls off asymptotically as the distance increases. Including electrodynamic corrections affects the peak values of the thermal radiation, making the radiative flux a vital observable for determining the magnetic charge.

\section{Numerical Analysis}\label{sec:5}

Here, we study the ISCOs of dyonic black holes in non-linear Electrodynamics. It should be noted that the ISCOs can be derived by setting to zero the second derivative of the effective potential. Thus, one can derive the dimensionless radius of the innermost stable circular geodesic orbits.

It should be emphasized that in order to have circular orbits, one must have $\bar{V}_{eff}=0$, and $\frac{d \bar{V}_{eff}}{dr}=0$.
In addition, innermost circular orbits correspond to the local minimum of the effective potential. Hence, the innermost (marginally) stable circular geodesic orbit is obtained from $\frac{d^2 \bar{V}_{eff}}{dr^2}=0$. Clearly, the ISCO corresponds to the transition between stable and unstable circular orbits \cite{Novikov:1973kta, Jiang:2021ajk}.

\begin{figure}
\centering
\includegraphics[width=0.5\textwidth]{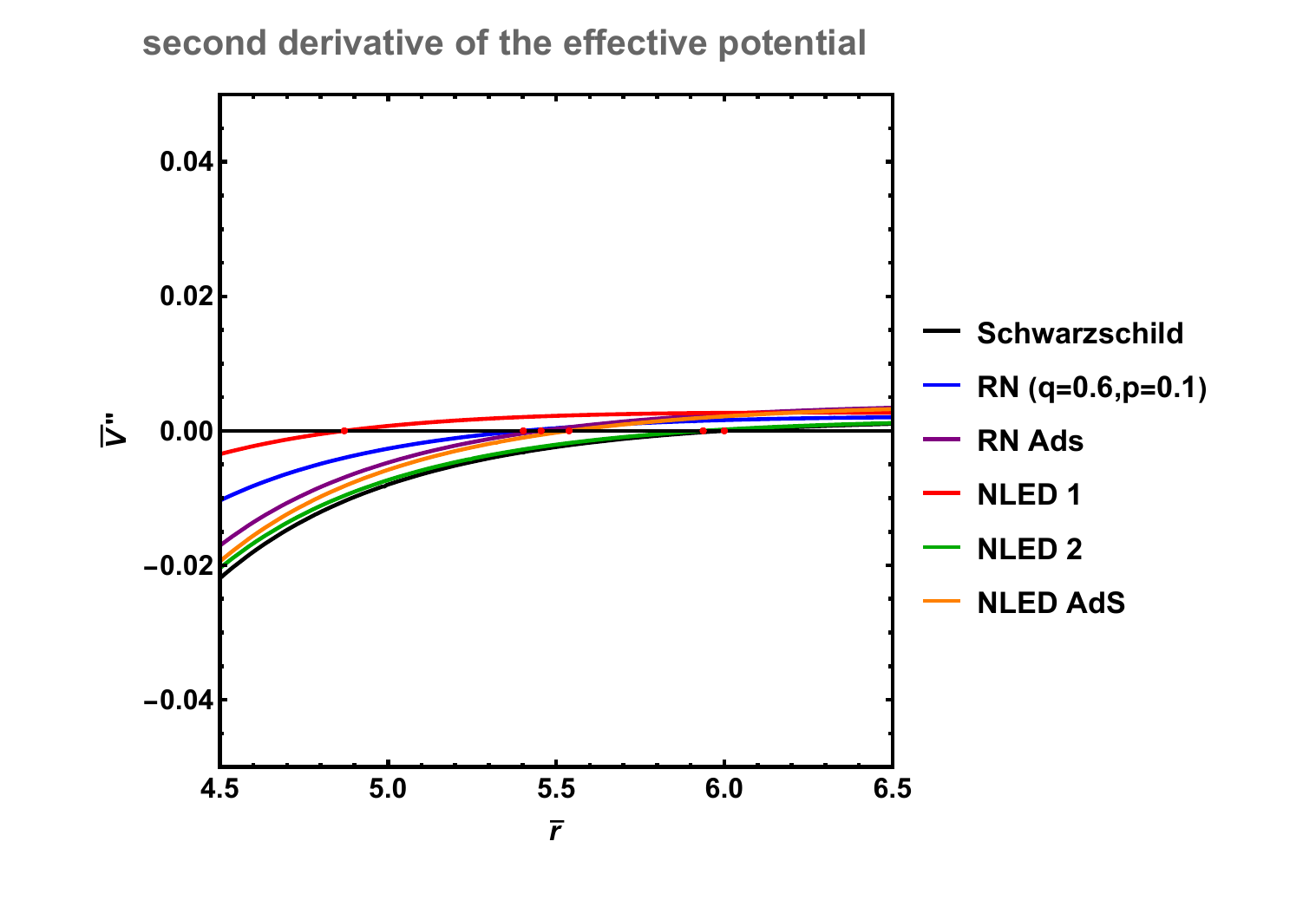}
\caption[figs]
{The second derivative of the effective potential of dyonic, Reissner-Nordstr\"om, and Schwarzschild black holes shows the locations of the ISCOs for the different cases.}
\label{fig17}
\end{figure}

In Fig. (\ref{fig17}), we explore the concept of ISCOs by illustrating the second derivative of the effective potential of Schwarzschild, Reissner-Nordstr\"om, and dyonic black holes with respect to $\bar{r}$. The figure illustrates the connection between the second derivative of the effective potential and the ISCOs, which are characterized by the solutions to the equation $\frac{d^{2}\bar{V}_{eff}}{d\bar{r}^{2}}=0$.

To show the validity of the method, we derived the standard ISCOs for Reissner-Nordstr\"om and Schwarzschild black holes in dimensionless form, which are completely consistent with the results \cite{Darwin:1959vua,Pugliese:2010ps,Chandrasekhar:1984siy}.

\begin{figure*}
\centering
\includegraphics[width=0.9\textwidth]{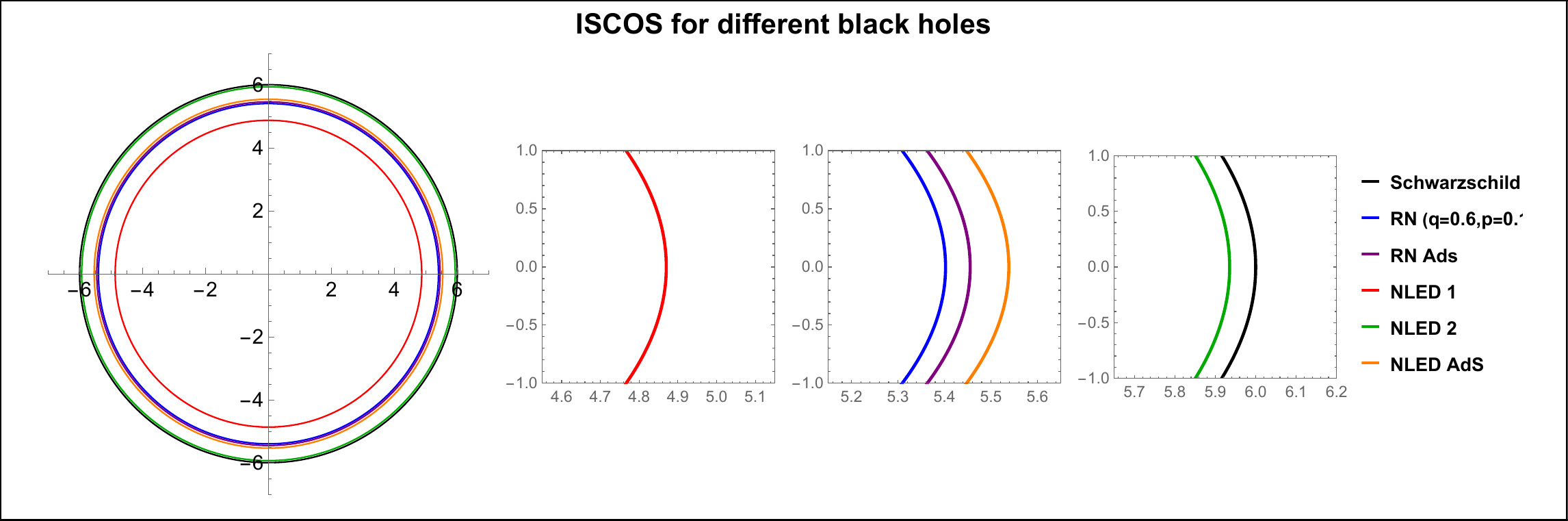}
\caption{Innermost Stable Circular Orbits of Schwarzschild, Reissner-Nordstr\"om, and dyonic black holes}
\label{fig18}
\end{figure*}

In Fig (\ref{fig18}), we have illustrated the innermost stable circular orbits of the Schwarzschild black hole, the Reissner-Nordstr\"om black hole, and dyonic black holes in Non-linear electrodynamics.

\begin{figure}
\centering
\includegraphics[width=7cm]{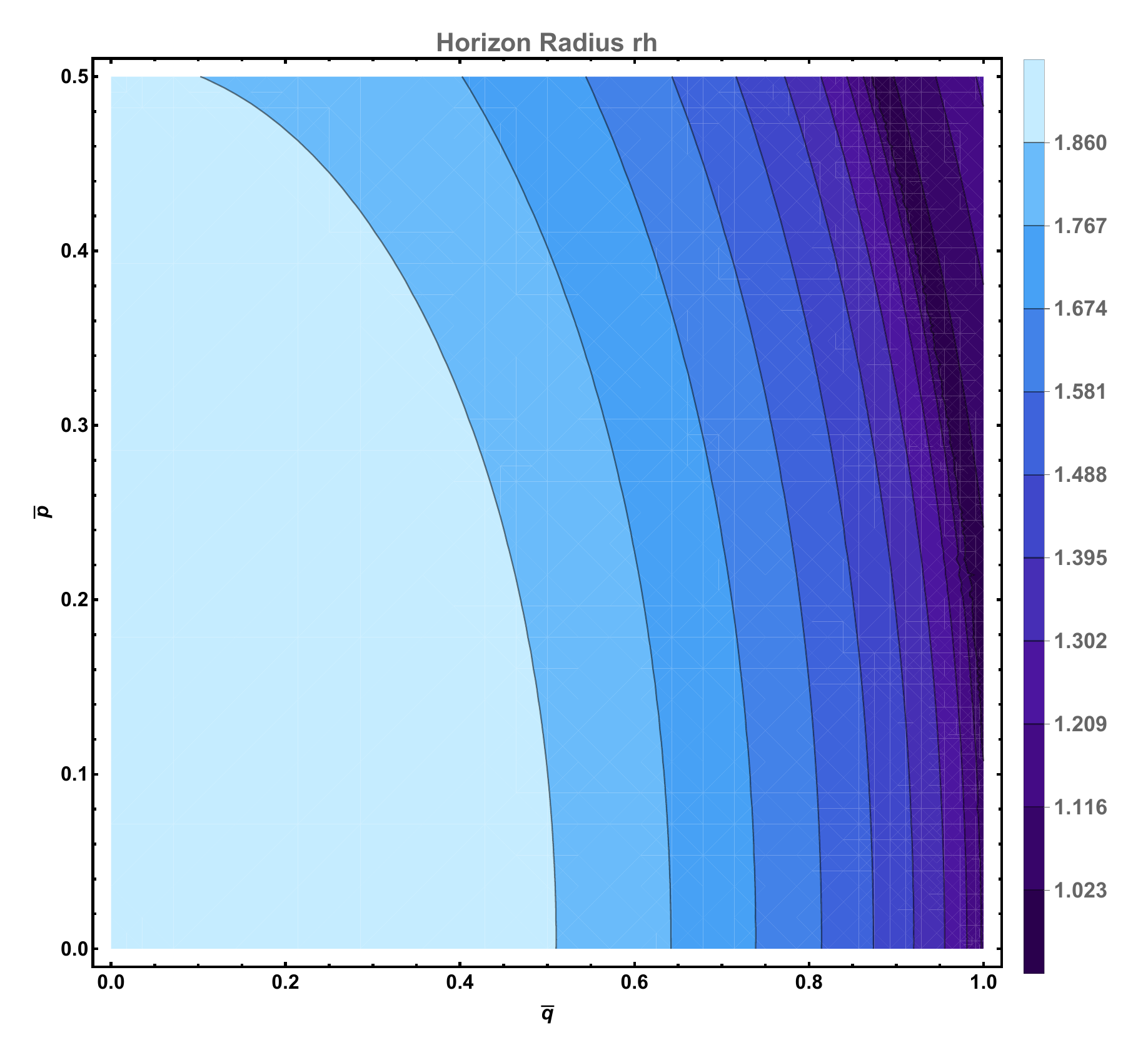}
\caption[figs]
{The event horizon radius for different values of $\bar{p}$ and $\bar{q}$ for dyonic black hole. The figur is plotted by considering the values $\bar{a}=0.037$, $\bar{b}=0.028$, $\bar{\Lambda}=0$.}
\label{fig19}
\end{figure}

\begin{figure}
\centering
\includegraphics[width=7cm]{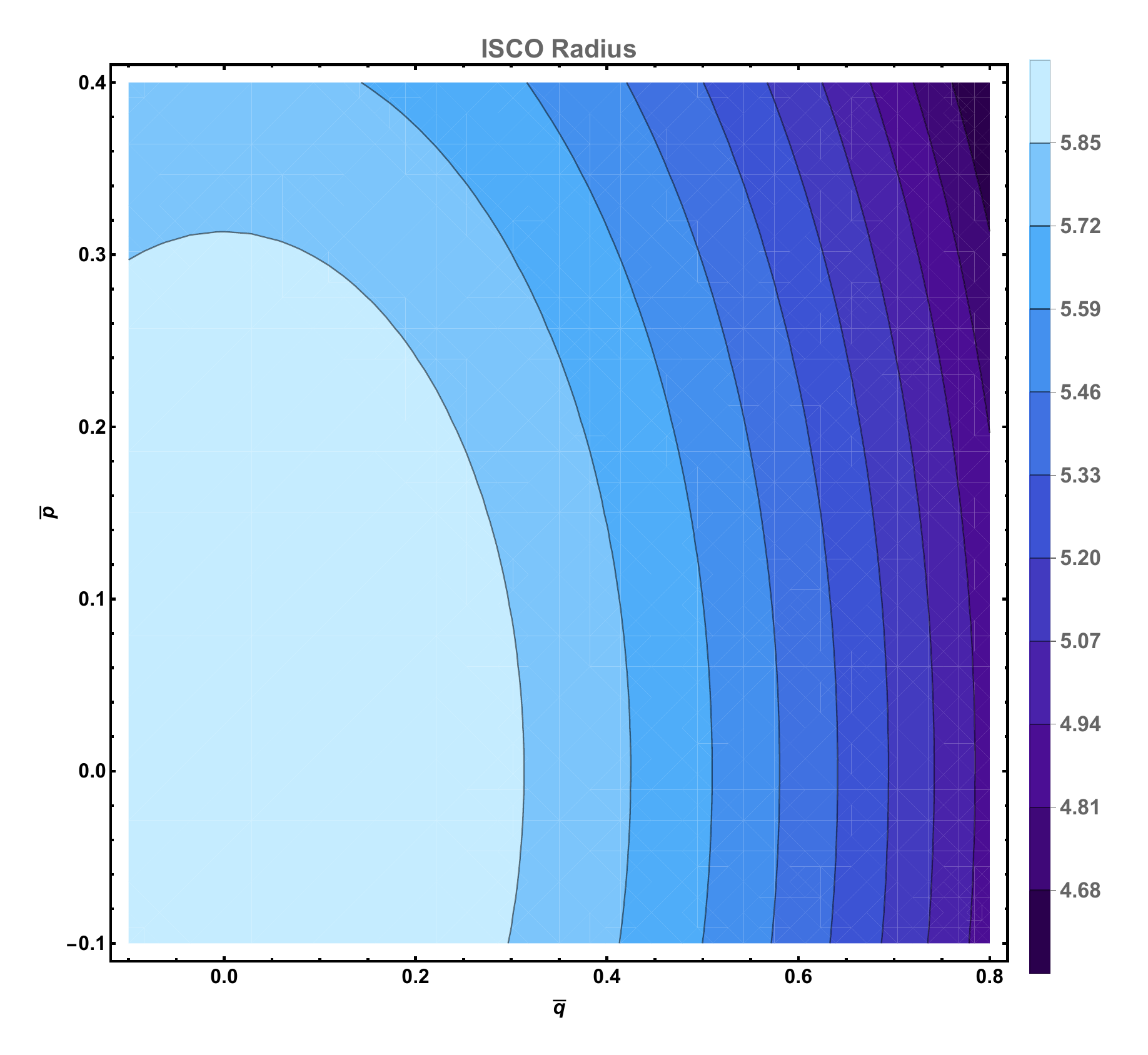}
\caption[figs]
{Innermost Stable Circular Orbits of a dyonic black hole for different values of $\bar{p}$ and $\bar{q}$ by considering $\bar{a}=0.037$, $\bar{b}=0.028$, $\bar{\Lambda}=0$.}
\label{fig20}
\end{figure}

The impact of magnetic charge $\bar{p}$ and electric charge $\bar{q}$ on event horizon radius $\bar{r}_{h}$ and ISCO radius $\bar{r}_{ISCO}$ in the case of a dyonic black hole is illustrated in a systematic way in Figs.~(\ref{fig19}) and (\ref{fig20}), respectively, with the help of fixed values of the NLED parameters $\bar{a}=0.037$ and $\bar{b}=0.028$ in the absence of cosmological constant ($\bar{\Lambda}=0$).

From Fig.~(\ref{fig19}), it is clear that the event horizon radius $\bar{r}_{h}$ is a strictly monotonically decreasing function of both electric and magnetic charge parameters. This is because an increase in the values of either $\bar{q}$ or $\bar{p}$ increases the repulsive electrostatic and magnetostatic forces inside the spacetime geometry.

The above-mentioned geometrical contraction directly affects the dynamics of external particles. Indeed, the analysis of Fig.~(\ref{fig20}) shows that the ISCOs radii $\bar{r}_{ISCO}$ decrease as the values of $\bar{p}$ and $\bar{q}$ increase, following the behavior of the horizon qualitatively. This occurs because the total additional charge reduces the effective gravitational force at large distances, allowing stable timelike circular orbits to exist at smaller $r$ before becoming radially unstable. Moreover, the comparison of Figs.~\eqref{fig20} and \eqref{fig18} shows that the ISCO radius always stays outside the horizon ($\bar{r}_{ISCO} > \bar{r}_{h}$), regardless of the chosen values of charge, guaranteeing the fact that the change of the circular orbit to the plunging one takes place within the bounds of our observable universe.

In the case of Magnetic black holes, the ISCO radius $\bar{r}_{ISCO}^{Mag}$ obeys $\bar{V}_{eff, Mag}''(\bar{r})=0$.  Applying the general condition~\eqref{eq:ISCO_condition} we obtain
\begin{eqnarray}\label{ISCO_Mag}
\bar{r}\,\bar{f}_{\text{Mag}}\,\bar{f}_{\text{Mag}}'' + 3\bar{f}_{\text{Mag}}\,\bar{f}_{\text{Mag}}' - 2\bar{r}\,(\bar{f}_{\text{Mag}}')^{2} = 0,
\end{eqnarray}

where the second derivative is

\begin{align}
\bar{f}_{\text{Mag}}''(\bar{r}) &= -\frac{4}{\bar{r}^{3}} - \frac{2\bar{\Lambda}}{3} + \frac{6\bar{p}^{2}}{\bar{r}^{4}} - \frac{84\bar{a}\bar{p}^{4}}{5\bar{r}^{8}}. \label{fpp_Mag}
\end{align}

Equation~\eqref{ISCO_Mag} is a high‑order algebraic equation in $\bar{r}$; it must be solved numerically for given parameters $\bar{\Lambda}$, $\bar{p}$, and $\bar{a}$.

\begin{figure}
\centering
\includegraphics[width=0.5\textwidth]{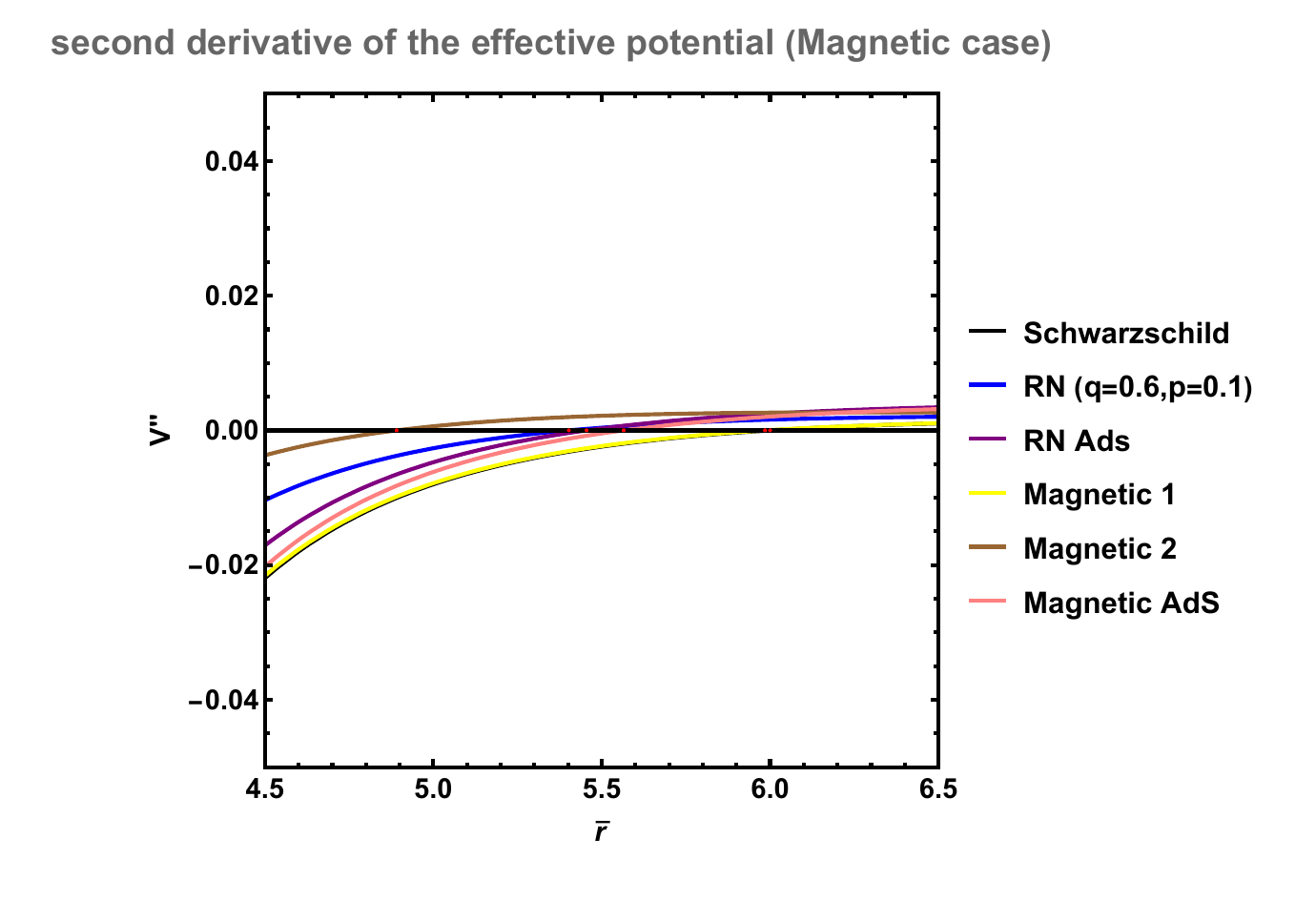}
\caption[figs]
{The second derivative of the effective potential of Magnetic, Reissner-Nordstr\"om, and Schwarzschild black holes shows the locations of the ISCOs for the different cases.}
\label{fig14}
\end{figure}

Spatial properties of the second derivative of the effective potential, $d^2\bar{V}_{\text{eff}}/d\bar{r}^2$, are presented in Fig.~(\ref{fig14}), where we consider three well-known solutions of Schwarzschild, Reissner-Nordstr\"om, and our proposed solution of a magnetic black hole. From the criteria of circular geodesic stability, it is clear that the roots of these graphs, at which $d^2\bar{V}_{\text{eff}}/d\bar{r}^2 = 0$, indicate the precise positions in space for the innermost stable circular orbits ($\bar{r}_{\text{ISCO}}$) of a test body in each of these three spaces. It can be easily noticed from the figure that, with the addition of the electric/magnetic charge parameter, the root of this curve moves towards smaller radii as compared to the vacuum Schwarzschild case ($\bar{r}_{\text{ISCO}} = 6$).

\begin{figure*}
\centering
\includegraphics[width=0.9\textwidth]{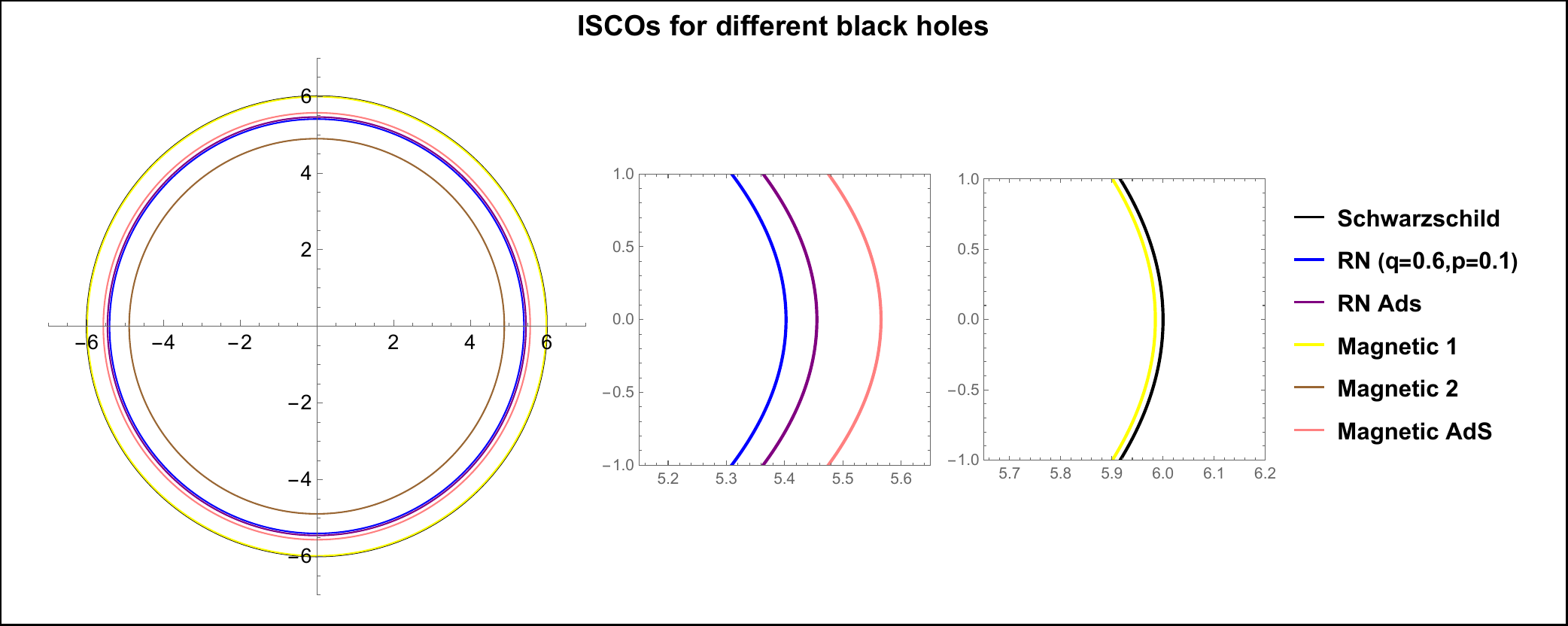}
\caption{Innermost Stable Circular Orbits of Magnetic, Schwarzschild, and Reissner-Nordstr\"om black holes.}
\label{fig15}
\end{figure*}

Fig. (\ref{fig15}) shows a comparison of the innermost stable circular orbits ($\bar{r}_{\text{ISCO}}$) in Magnetic, Schwarzschild, and Reissner-Nordstr\"om black hole spacetimes. This figure clearly shows that the orbit structure depends on the source parameters. Although the Schwarzschild black hole holds its rigid classical shape with a circular boundary at $\bar{r}_{\text{ISCO}} = 6$, the inclusion of electric charge and the non-linear magnetic field reduces the circular boundary to an inner region due to their respective effects. The non-linear magnetic field produces a distinctive physical structure in which test particles can move in stable circular orbits closer to the central object.

\begin{table*}
\caption{\label{tab:table1}Locations of the event horizons and the ISCOs in dimensionless conditions for a Schwarzschild black hole, a Reissner-Nordstr\"om black hole, and a dyonic black hole in Non-linear electrodynamics.}
\begin{tabular}{ccccc}\hline
Black Hole & Event horizon &Innermost Stable circular orbits \\ \hline
Schwarzschild &  2 & 6 \\ \hline
Reissner-Nordstr\"om (RN) & 1.79 & 5.40 \\ \hline
Reissner-Nordstr\"om (RN AdS) & 1.94 & 5.45 \\ \hline
NLED 1 & 1.59 & 4.87\\ \hline
NLED 2 &1.97 & 5.93 \\ \hline
NLED AdS &1.98 & 5.53 \\ \hline
Magnetic 1 &1.99 &5.98\\ \hline
Magnetic 2 &1.60 &4.89\\ \hline
Magnetic AdS &1.99 &5.56\\ \hline
\end{tabular}
\end{table*}

In Table \ref{tab:table1}, we present the locations of the event horizons and ISCOs in dimensionless form for various black hole models, including the Schwarzschild black hole, Reissner-Nordstr\"om black hole, and a dyonic black hole. 

\subsection{Observable Signatures and Astrophysical Implications}\label{ObS}

Now that we have discussed the purely geometric/geodesic aspects of the spacetime models under consideration, let us turn our attention to their concrete physical manifestations in the form of astronomy. We can identify signatures of the modification due to non-linear electrodynamics by studying the motion of matter around the black hole in the equatorial plane. This process will allow us to differentiate between new black hole models and those of general relativity using astronomical observations.

\subsubsection{Thermal Spectrum and Radiative Efficiency}

Here, we can study the continuum thermal spectrum and the radiative efficiency of a thin accretion disk around a dyonic black hole in NLED. The NLED correction to the gravitational potential changes the innermost stable circular orbit (ISCO). The ISCO affects the disk temperature, spectrum, and radiative efficiency. Such effects serve as observational evidence for distinguishing the NLED black hole from its other counterparts (the Reissner-Nordstr\"om and Schwarzschild black holes).

The continuum thermal spectrum is due to multicolor blackbody radiation emitted from various radii of the accretion disk. Each annulus emits a blackbody spectrum of temperature $T(r)$, and the total spectrum is the sum over all annuli \cite{Shakura:1972te, Novikov:1973kta}.

The thermal spectra arise from multicolor blackbody radiation emitted from different radial regions of the accretion disk. Although the geometric configuration and the energy flux of the relativistic thin disk have been determined by Shakura and Sunyaev \cite{Shakura:1972te} and Novikov and Thorne \cite{Novikov:1973kta} (as well as Page and Thorne \cite{Page:1974he}), each region radiates as a blackbody at temperature $T(r)$. The overall radiation spectrum is calculated by integrating the blackbody radiation from all the stable annular regions of the disk starting from the ISCO outward \cite{Mitsuda:1984nv}.

From Eq. (\ref{Eq19}) for radiative flux $F(r)$, the effective temperature using Stefan-Bolt can be obtained,

\begin{eqnarray}
T_{eff}(r) = [\frac{F(r)}{\sigma}]^{1/4}.
\end{eqnarray}

\begin{figure}
\centering
\includegraphics[width=0.5\textwidth]{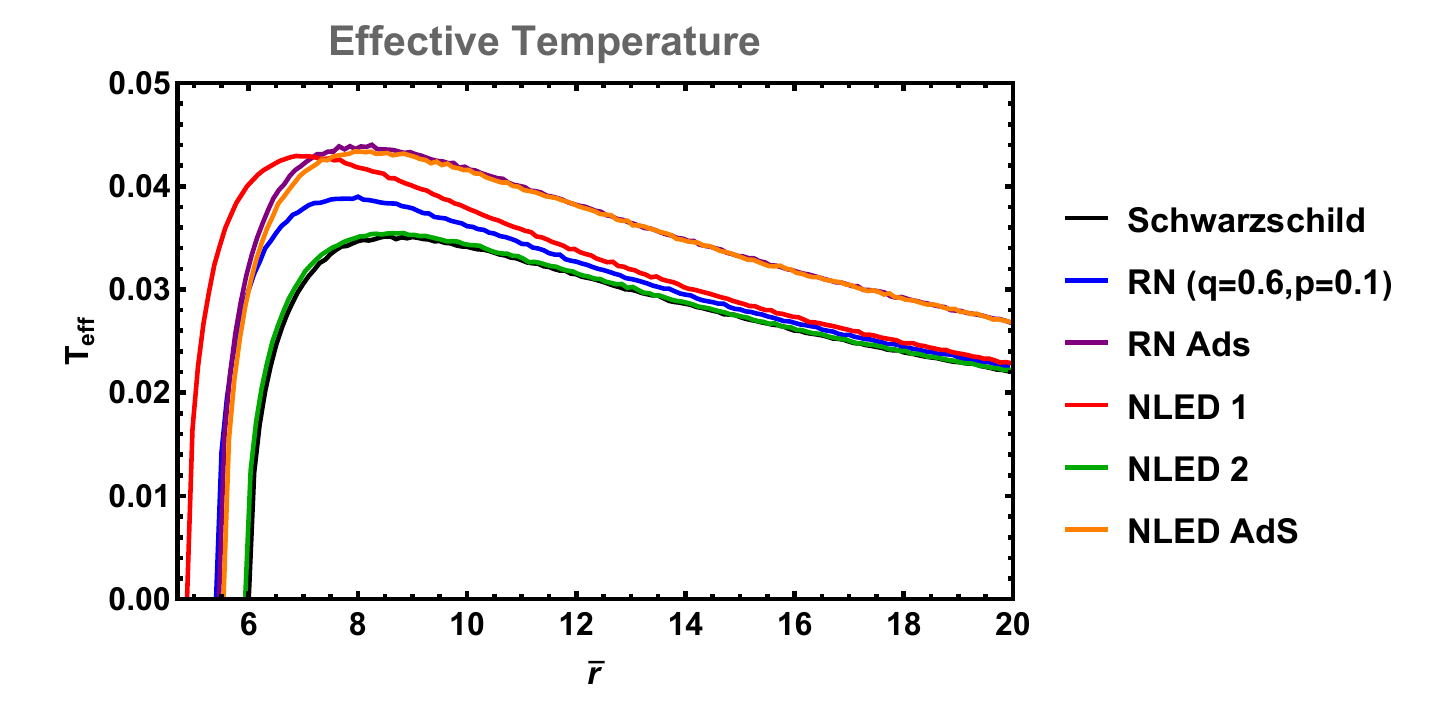}
\caption{The graphs show the effective temperature for Schwarzschild, Reissner-Nordstr\"om, and dyonic black holes.}
\label{fig210}
\end{figure}

\begin{figure*}
\centering
\includegraphics[width=0.9\textwidth]{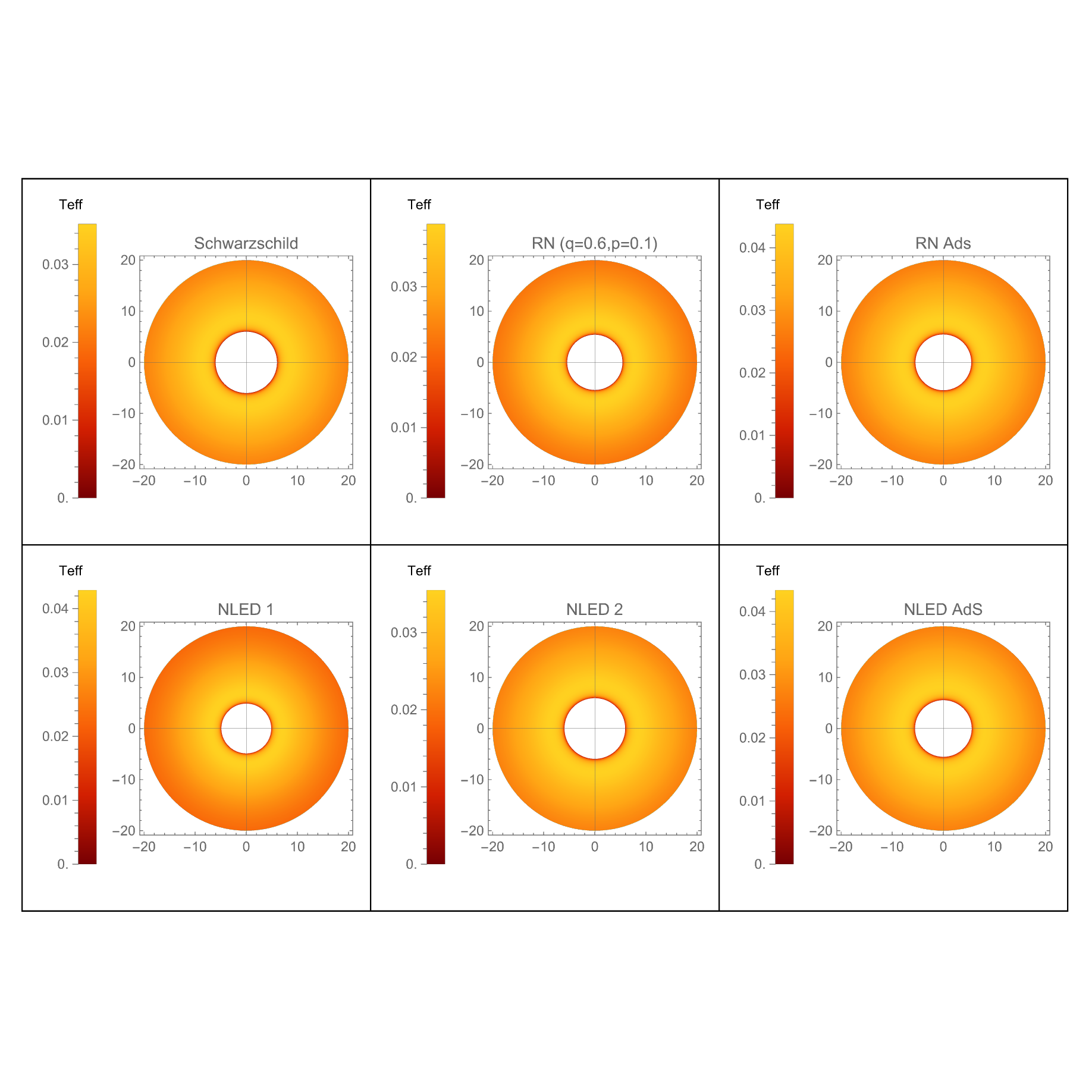}
\caption{Effective temperature for Schwarzschild, Reissner-Nordstr\"om, and dyonic black holes.}
\label{fig21}
\end{figure*}

In Figs. (\ref{fig210}) and (\ref{fig21}), we display the graph of the temperature profile $T(r)$ for a thin accretion disc in NLEDs around Schwarzschild, RN, and the new dyonic black holes as functions of the dimensionless radial coordinate $\bar{r}$. While Fig. (\ref{fig210}) shows an overlapped graph of these temperatures, the multi-paneled Fig. (\ref{fig21}) is used to better discern the effect of geometry on the temperatures.

As predicted by the Page-Thorne theories, all three graphs start from zero at the inner boundary radius ($\bar{r}_{ISCO}$) of each configuration due to zero torque applied at that point. After the sudden increase from zero at the ISCO radius, the temperature quickly rises to reach a sharp peak beyond the ISCO radius before falling to the universal asymptotic power-law form of $T(r) \propto r^{-3/4}$ \cite{Shakura:1972te}. As per the established concept of a thin accretion disk, the graphs of the temperature distributions shown in Figs. (\ref{fig21}) and (\ref{fig10Mg}) have been presented only outside the innermost stable circular orbit ($r \ge r_{ISCO}$). Inside the ISCO ($r < r_{ISCO}$), the disk disappears because of the falling of test particles towards the black hole and the absence of thermal emission from there.

In addition, the plots indicate that the thermal properties are highly sensitive to the charge distribution in the compact star. The presence of the ordinary electric charge in the RN metric, and especially the presence of both the electric and the non-linear magnetic charge in the case of the dyonic metric, causes the thermal peak to move to smaller radial positions and greatly increase the peak magnitude compared to the Schwarzschild vacuum background. This enhancement arises from the closer proximity of the ISCOs, allowing the accretion disk to approach the strong-field gravity region and release more gravitational potential energy as heat.

Concerning the efficiency of the radiative process $\eta$, one should note that the fraction of the rest-mass energy of accreted matter that goes into radiation. An important parameter related to the efficiency of the process of accretion is the efficiency of radiative processes $\eta$, which is the fraction of the rest mass energy of accreted matter that gets converted to radiation before matter falls into the black hole. For a particle starting from infinity and reaching the ISCO, the efficiency is determined by the specific energy at ISCO, $\bar{E}_{ISCO} = E_{ISCO}/m_{rest}$, as follows \cite{Novikov:1973kta, Page:1974he}:

For a particle falling from infinity to ISCO,
\begin{eqnarray}
\eta = 1 - \bar{E}_{ISCO}.
\end{eqnarray}

For the Schwarzschild black hole $\bar{E}_{ISCO} = \sqrt{8/9} \approx 0.9428$, hence $\eta \approx 0.0572$ \cite{Bardeen:1972fi, Thorne:1974ve}. For the Reissner–Nordstr\"om black holes, the efficiency is usually larger since the ISCO gets closer to the black hole. NLED effects may either enhance or diminish $\eta$ depending on the coupling sign and strength.

By using $\bar{E}(\bar{r})$ from Eq. (\ref{eq:E_circ}) and the ISCO radii $\bar{r}_{ISCO}$ obtained in Section~(\ref{sec:5}), we estimate $\bar{E}_{\text{ISCO}}$ for each considered type of black hole. The values are summarized in Table~(\ref{tab:efficiency}). For the NLED models, the efficiency can be as high as $\sim 10\%$ above the Schwarzschild value, which would certainly affect the emission luminosity. These discrepancies give another way to check NLED theories against X-ray observations of quasars.
\begin{figure}
\centering
\includegraphics[width=0.5\textwidth]{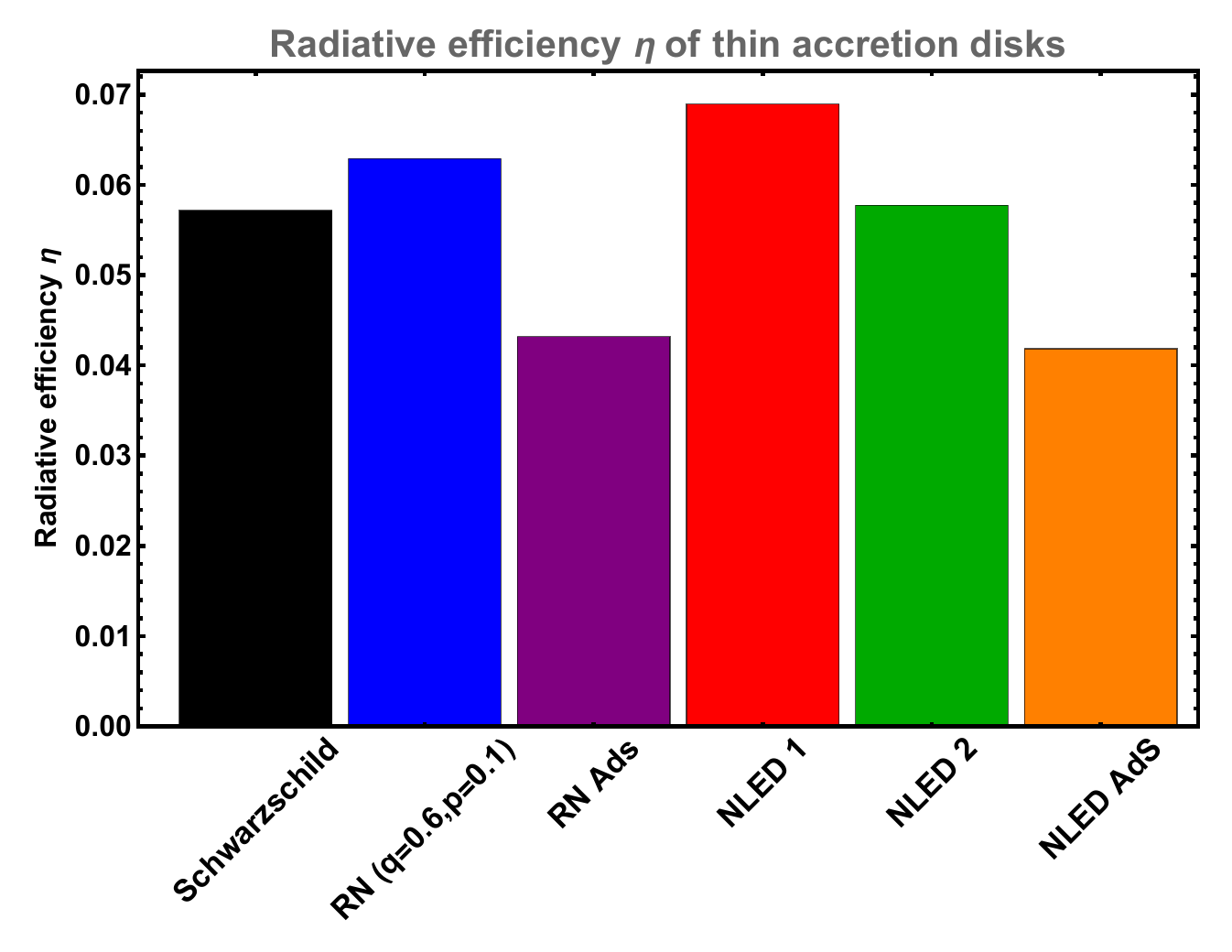}
\caption{The graphs show the radiative efficiency for Schwarzschild, Reissner-Nordstr\"om, and dyonic black holes.}
\label{fig11}
\end{figure}

The radiative efficiency of various types of black hole solutions has been shown in Fig.(\ref{fig11}). As shown in the graph, the effect of the charge parameters is evident in the efficiency curves. The main point is that the presence of the electric and magnetic charges of the dyonic black hole solution impacts significantly on the efficiency curves and their peak efficiency as compared to the usual Schwarzschild and Reissner-Nordstr\"om solutions. Increasing the charge parameter decreases the distance between the ISCO and the horizon, thus increasing efficiency.

\begin{table}
\centering
\caption{Radiative efficiency $\eta = 1 - \bar{E}_{\text{ISCO}}$ for each black hole model.}
\label{tab:efficiency}
\begin{tabular}{lccc}
\hline
Model & $\bar{r}_{\text{ISCO}}$ & $\bar{E}_{ISCO}$ & $\eta$ \\ \hline
Schwarzschild         & 6.000 & 0.943 & 0.0570 \\
RN ($q=0.8,p=0.1$)    & 5.402 & 0.937 & 0.0629 \\
RN ($q=0.3,p=0.1$)AdS & 5.455 & 0.957 & 0.0430 \\
NLED 1                & 4.870 & 0.931 & 0.0690 \\
NLED 2                & 5.935 & 0.942 & 0.0580 \\
NLED AdS              & 5.538 & 0.958 & 0.0420 \\ \hline
\end{tabular}
\end{table}

\subsubsection{The Photon Sphere and Black Hole Shadow}

Another key observable in the regime of strong gravitational fields is the black hole shadow: the dark area on an observer's celestial sphere created by the black hole. In the case of a static and spherically symmetric spacetime, the black hole shadow is circular, and the radius of the shadow is defined by the unstable photon orbits (photon sphere). The effective potential of the null geodesics ($\epsilon = 0$) is
\begin{eqnarray}
\bar{V}_{\text{eff, photon}}(\bar{r}) = \frac{\bar{f}(\bar{r})}{\bar{r}^{2}} \bar{L}^{2},
\end{eqnarray}

where $\bar{L}$ denotes the conserved angular momentum. 

The radius of the photon sphere $\bar{r}_{\text{ph}}$ does not have an analytical expression due to the presence of non-linear electrodynamic effects in the metric. So, $\bar{r}_{\text{ph}}$ is calculated using numerical methods as the largest real root greater than the event horizon ($\bar{r}_{\text{ph}} > \bar{r}_h$), which solves the equation $d\bar{V}_{\text{eff, photon}}/d\bar{r} = 0$. The above equation takes the following algebraic form:

\begin{eqnarray}
2 \bar{f}(\bar{r}_{\text{ph}}) - \bar{r}_{\text{ph}} \bar{f}'(\bar{r}_{\text{ph}}) = 0.
\end{eqnarray}

The maximum root $\bar{r}_{\text{ph}} > \bar{r}_{h}$ of this equation corresponds to the radius of the photon sphere. For a Schwarzschild black hole, we have that $\bar{r}_{\text{ph}} = 3$ \cite{Synge:1966okc, Bardeen:1972fi}, whereas for a Reissner-Nordstr\"om solution, $\bar{r}_{\text{ph}} < 3$ and diminishes even more with the increase of charge.

After obtaining the correct value of $\bar{r}_{\text{ph}}$ for any particular set of parameter couplings $(\bar{a}, \bar{b}, \bar{q}, \bar{p}, \bar{\Lambda})$ through our numerical approach, we then plug it into the formula for the critical impact parameter, which determines the apparent radius of the circular black hole shadow for the asymptotic observer.

The black hole shadow radius is

\begin{eqnarray}
\bar{b}_{\text{crit}} = \frac{\bar{r}_{\text{ph}}}{\sqrt{\bar{f}(\bar{r}_{\text{ph}})}}.
\end{eqnarray}

As far as the celestial coordinates $(\alpha, \beta)$ are concerned, relative to an asymptotic observer at an inclination angle $i$, the appearance of the shadow is defined through projection of the momentum vectors of the null geodesics on the observer’s sky coordinate system \cite{Bardeen:1972fi}. In the case of spherical symmetry, the impact parameters take a very simple form:
\begin{equation}
\bar{\alpha} = \lim_{\bar{r}_0 \to \infty} \left( -\bar{r}_0^2 \sin i \frac{d\phi}{d\bar{t}} \right) = -\frac{\bar{L}}{\sin i},
\label{eq:alpha_celestial}
\end{equation}
\begin{equation}
\bar{\beta} = \lim_{\bar{r}_0 \to \infty} \left( \bar{r}_0^2 \frac{d\theta}{d\bar{t}} \right) = \pm \sqrt{\bar{\mathcal{Q}} + \bar{L}^2 \cot^2 i}.
\label{eq:beta_celestial}
\end{equation}

In this case, $\bar{L}$ refers to the angular momentum without dimensions, $\bar{\mathcal{Q}}$ refers to the dimensionless Carter constant, and $r_0$ stands for the radius of the far-away observer. If an observer is located at the equatorial plane with the inclination angle of $i=\pi/2$, then celestial coordinates become independent. This yields the standard parameterization constraint $\bar{\alpha}^2 + \bar{\beta}^2 = \bar{L}^2 + \bar{\mathcal{Q}} = \bar{b}_{\text{crit}}^2$.

\begin{figure*}
\centering
\includegraphics[width=0.9\textwidth]{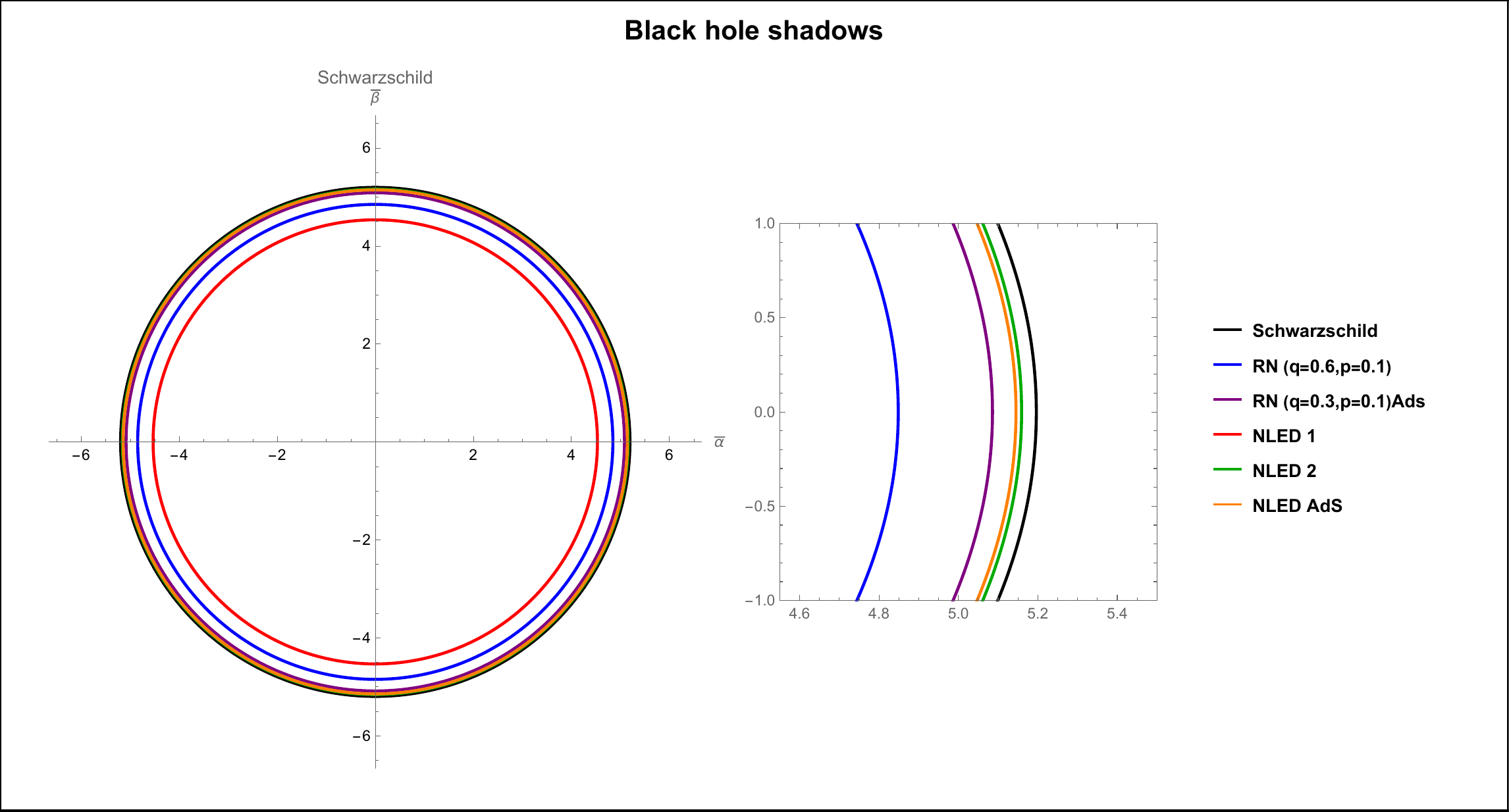}
\caption{Celestial plane shadows (circles) for the different black hole models. The radius of each circle is the critical impact parameter $\bar{b}_{\text{crit}}$. The Schwarzschild shadow is shown as a reference.}
\label{fig:shadows23}
\end{figure*}

The critical impact parameter $\bar{b}_{\text{crit}}$ and the corresponding photon sphere radius $\bar{r}_{\text{ph}}$ are calculated numerically for all the black hole models using the interpolated metric component $\bar{f}(\bar{r})$. As for the basic Schwarzschild case, we have numerically obtained the analytical solutions $\bar{r}_{\text{ph}}=3$ and $\bar{b}_{\text{crit}} = 3\sqrt{3}\approx5.196$, thereby demonstrating the reliability of the calculation algorithm \cite{Synge:1966okc, Bardeen:1972fi}. Since the photon spheres have been modified owing to non-linear electrodynamics, the shadow contours of the NLED black holes show clear deviations from the Schwarzschild and RN metrics. In particular, these deviations in geometry are significantly seen when the coupling constants of the NLED black holes ($a,b$) and the cosmological constant $\Lambda$ are high. The shadow contours for different black holes in this case are shown in Fig. (\ref{fig:shadows23}).

\begin{table}
\centering
\caption{Photon sphere radius $\bar{r}_{\text{ph}}$ and shadow radius $b_{\text{crit}}/M$ for the black hole models considered.}
\label{tab:shadows}
\begin{tabular}{lcc}
\hline
Model & $\bar{r}_{\text{ph}}$ & $\bar{b}_{\text{crit}}$ \\ \hline
Schwarzschild & 3.000 & 5.196 \\
RN ($q=0.8,p=0.1$) & 2.729 & 4.849 \\
RN ($q=0.3,p=0.1$)AdS & 2.932 & 5.086 \\
NLED 1 & 2.475 & 4.534 \\
NLED 2 & 2.971 & 5.159 \\
NLED AdS & 2.978 & 5.145 \\ \hline
\end{tabular}
\end{table}

The shadow of the supermassive black hole present at the Galactic center, Sgr A*, has been resolved through the use of the EHT, resulting in a measurement of the angular size of the black hole shadow \cite{EventHorizonTelescope:2022xqj, EventHorizonTelescope:2022wkp}.
For our numerical comparison purposes, we will take into consideration a black hole physical mass of $m_{\text{phys}} = 4.0\times10^{6}\,M_{\odot}$ \cite{EventHorizonTelescope:2022xqj} at an effective baseline scaling distance of  $D_{\text{phys}} \approx 7.92\;\text{kpc}$ \cite{Gravity:2019nxk}, which corresponds to the physical scaling used for our geometric transformation from dimensionless to physical units.

Using the dimensionless metric function $\bar{f}(\bar{r})$, we compute the dimensionless critical impact parameter $\bar{b}_{\text{crit}} = \bar{r}_{\text{ph}}/\sqrt{\bar{f}(\bar{r}_{\text{ph}})}$ for each black hole model (see Table~\ref{tab:shadows}). 

The physical angular diameter in microarcseconds ($\mu\text{as}$) is then recovered via the standard geometric projection:

\begin{equation}
\theta = 2 \bar{b}_{\text{crit}} \frac{G m_{\text{phys}}}{c^2 D_{\text{phys}}} \times \mathcal{C},
\label{eq:theta_phys}
\end{equation}

where $G$ is the gravitational constant, $c$ is the speed of light, and $\mathcal{C} = 206265 \times 10^6$ represents the conversion factor from radians to microarcseconds.

 The calculated angular sizes are shown in Table~\ref{tab:sgra_comparison}. The deviation of shadow size in the case of Reissner-Nordstr\"om and NLED models arises because of the altered gravitational potential.

In order to quantify the goodness of fit of each theoretical model against the EHT observations of Sgr A*, we use the standard $\chi^{2}$ goodness-of-fit formula:

\begin{equation}
\chi^{2} = \left(\frac{\theta_{\text{model}} - \theta_{\text{obs}}}{\sigma_{\text{obs}}}\right)^{2},
\label{eq:chi2_definition}
\end{equation}

where $\theta_{\text{obs}} = 48.7\;\mu\text{as}$ and $\sigma_{\text{obs}} = 7.0\;\mu\text{as}$ correspond to the EHT Sgr A* observational data. The models for which $\chi^{2} < 1$ indicate excellent agreement within the $1\sigma$ experimental error.

\begin{figure}
\centering
\includegraphics[width=0.5\textwidth]{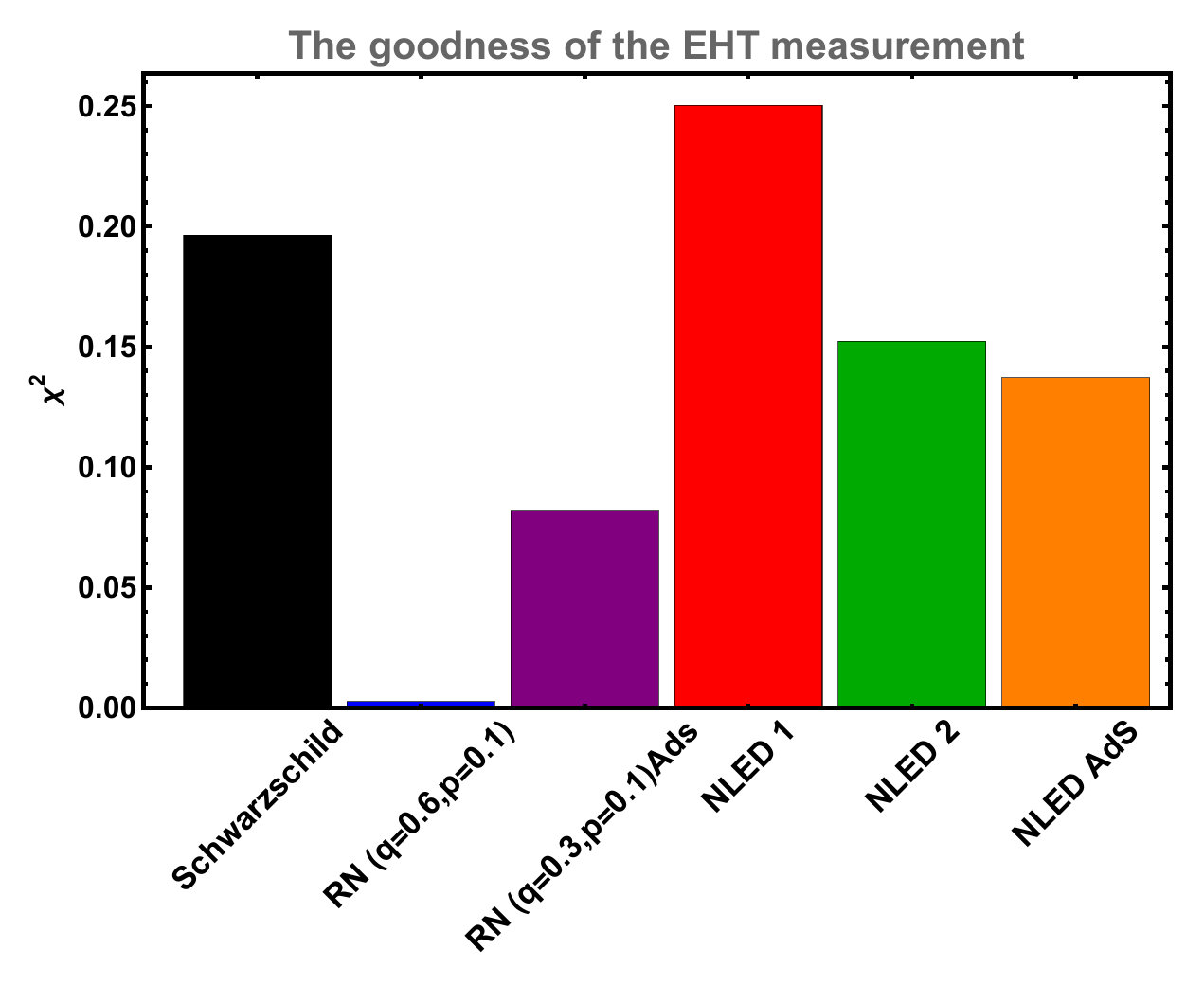}
\caption{The $\chi^2$ of angular diameters of the black hole shadow for the different models.}
\label{x2}
\end{figure}

Fig. (\ref{x2}) depicts the values of $\chi^{2}$ obtained from the angular diameter of the black hole shadow in the theoretical models under consideration. It is clear that such a comparative approach is useful for understanding how the given theory can fit into the observational constraints. The point at which the value of the $\chi^{2}$ distribution is minimal corresponds to the point in the parameter space where the angular diameter coincides with the observed data.

\begin{figure}
\centering
\includegraphics[width=0.5\textwidth]{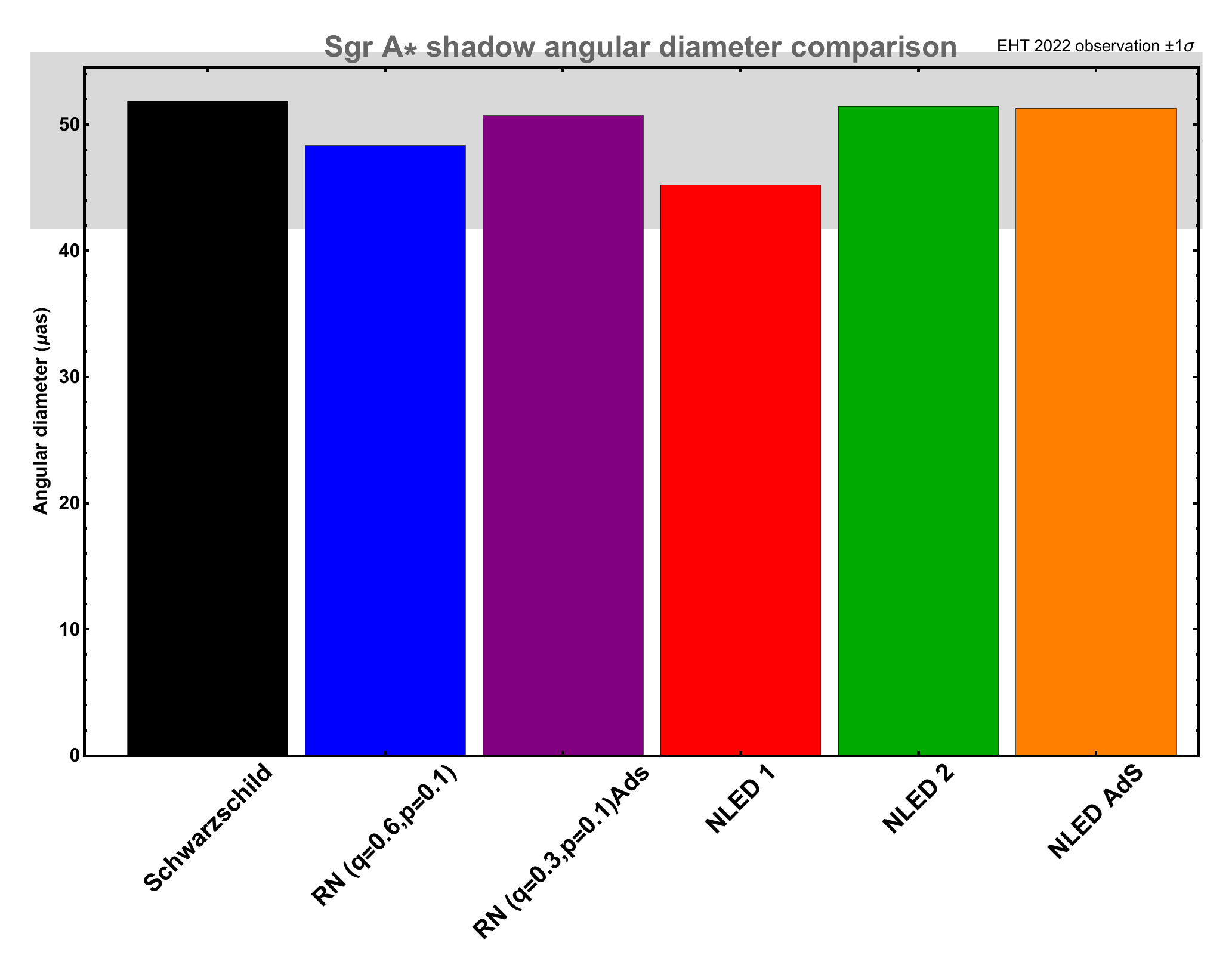}
\caption{Angular diameters of the black hole shadow for the different models, compared with the EHT observation of Sgr A* (dashed line) and its $1\sigma$ uncertainty (gray band). Models with bars intersecting the gray band are consistent with the data at the $1\sigma$ level.}
\label{fig:sgrA_comparison}
\end{figure}

Fig. (\ref{fig:sgrA_comparison}) shows the agreement of theoretical angular diameters of the black hole shadow with the result of the EHT observations. As can be seen from Fig.  (\ref{fig:sgrA_comparison}), the dashed horizontal line with a gray band around it is a value obtained from EHT with the $1\sigma$ error band. All NLED models discussed here have the shadow diameter in the $2\sigma$ region of the EHT measurement, while the best model is the one with the parameters $(a,b,q,p,\Lambda)$ at $\theta = 48.7\;\mu\text{as}$. Thus, NLED modifications cannot be ruled out based on the current EHT measurements of the Sgr A* shadow. More accurate measurements in the future (e.g., next-generation EHT) will help to select the models and put restrictions on the NLED coupling constants $a$ and $b$.

\begin{table}
\centering
\caption{Angular diameter of the black hole shadow for each model, compared with the EHT measurement for Sgr A*. The $\chi^2$ value indicates the goodness of fit.}
\label{tab:sgra_comparison}
\begin{tabular}{lccc}
\hline
Model & $\bar{b}_{\text{crit}}$ & $\theta$ ($\mu$as) & $\chi^2$ \\ \hline
Schwarzschild         & 5.196 & 51.80 & 0.1964 \\
RN ($q=0.8,p=0.1$)    & 4.849 & 48.34 & 0.0026 \\
RN ($q=0.3,p=0.1$)AdS & 5.086 & 50.70 & 0.0817 \\
NLED 1                & 4.534 & 45.20 & 0.2503 \\
NLED 2                & 5.159 & 51.43 & 0.1523 \\
NLED AdS              & 5.145 & 51.29 & 0.1371 \\ \hline
\end{tabular}
\end{table}

\subsection{The Magnetic Black Hole}

We summarize the results of the remaining magnetic black hole models in this section. Effective temperature can be derived based on the Stefan-Boltzmann law, $T_{\text{eff}}(\bar{r}) = [F_{\text{Mag}}(\bar{r})/\sigma]^{1/4}$. The remaining observables (radiative efficiency, photon sphere, black hole shadow) are obtained via substitution of $\bar{f}_{\text{Mag}}$ and its derivatives into the general formulae given in Sec. (\ref{sec:5}--\ref{ObS}). The magnetic solution thus offers those figures.

\begin{figure}
\centering
\includegraphics[width=0.5\textwidth]{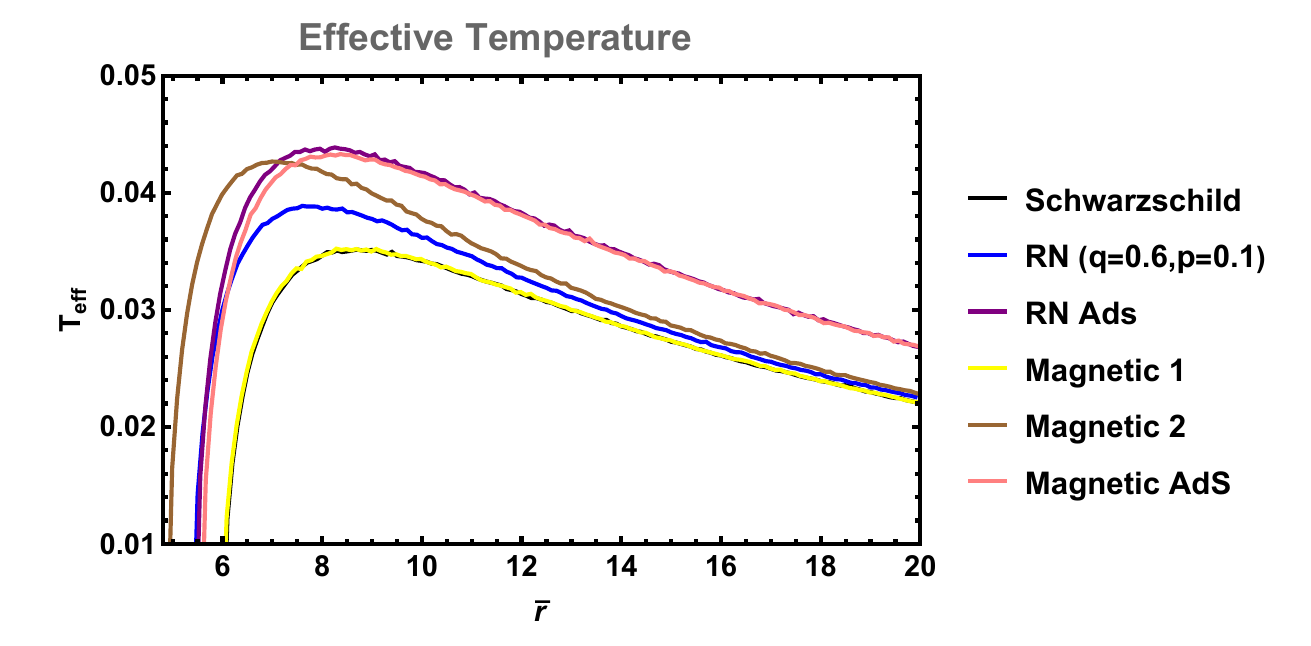}
\caption{The graphs show the effective temperature for Schwarzschild, Reissner-Nordstr\"om, and Magnetic black holes.}
\label{fig260Mg}
\end{figure}

\begin{figure*}
\centering
\includegraphics[width=0.9\textwidth]{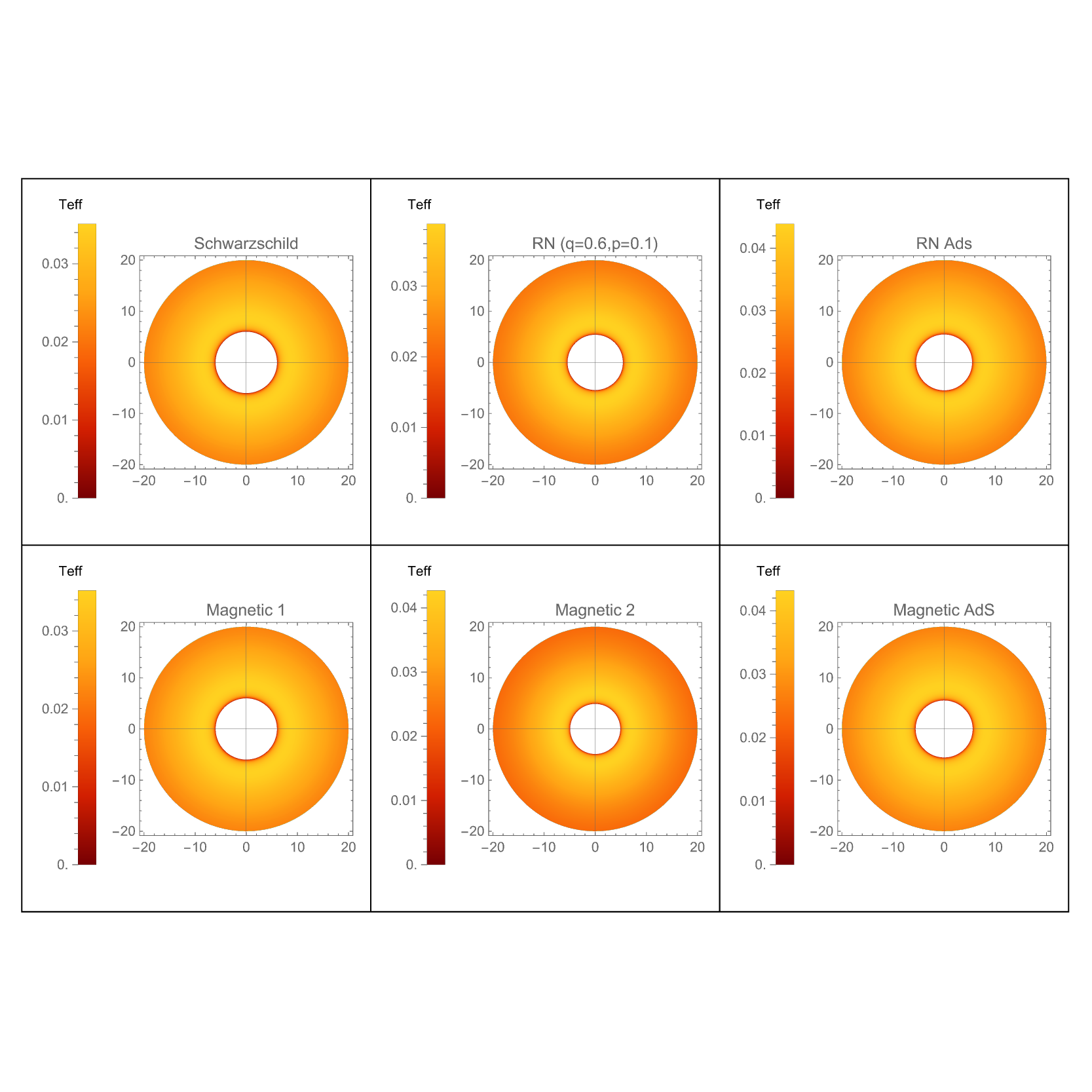}
\caption{Effective temperature for Schwarzschild, Reissner-Nordstr\"om, and Magnetic black holes.}
\label{fig10Mg}
\end{figure*}

In Figs. (\ref{fig260Mg}) and (\ref{fig10Mg}), the local effective temperature $T_{\text{eff}}(\bar{r})$ of the equatorial plane of the thin accretion disk is presented. Fig. (\ref{fig260Mg}) shows the comparative overlay of the temperature profiles, whereas Fig. (\ref{fig10Mg}) isolates these behaviors in separate subplots. In accordance with the boundary conditions of the relativistic thin disks, all curves are equal to zero at the respective points of the innermost stable circular orbit ($r=\bar{r}_{\text{ISCO}}$) where the internal torque equals zero.

Thermal behavior depends strongly on the field parameters. Specifically, the Magnetic 2 field configuration shifts the thermal maximum toward the center and significantly increases its value compared with the Schwarzschild case. This arises from the compression of the ISCO region; since stronger gravity liberates more gravitational potential energy as thermal radiation, the peak temperatures are higher in this configuration. The high-temperature peak could provide a distinctive thermodynamic feature that X-ray continuum spectroscopy could confirm.

\begin{figure}
\centering
\includegraphics[width=0.5\textwidth]{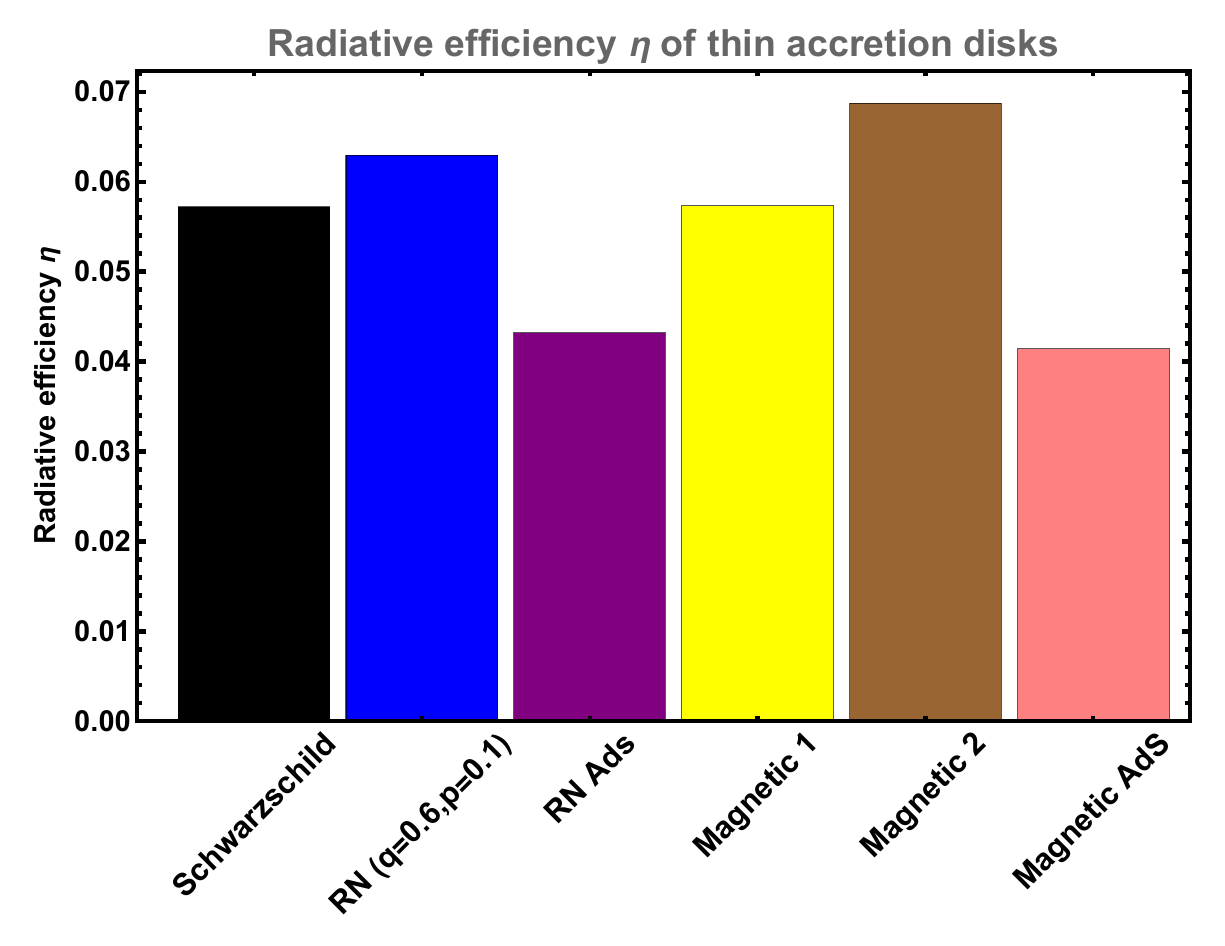}
\caption{The graphs show the radiative efficiency for Schwarzschild, Reissner-Nordstr\"om, and dyonic black holes.}
\label{fig11Mag}
\end{figure}

Table (\ref{tab:efficiency-magnetic}) shows the values for the energy at the innermost stable circular orbit ($\bar{E}_{\text{ISCO}}$) and the corresponding efficiencies ($\eta$), whereas Fig. (\ref{fig11Mag}) shows the trends in efficiencies within the different setups. The uncharged, static Schwarzschild solution serves as a benchmark and yields an efficiency of $\eta \approx 0.0570$. For purely non-linear magnetic cases, we note that the impact of the metric change on efficiency depends on the coupling constants and space-curvature properties. 

Magnetic 1 represents a situation in which the metric perturbation slightly pushes the stability boundary inward ($\bar{r}_{\text{ISCO}} = 5.985$) and yields a modest efficiency increase up to $5.73\%$. However, the more powerful magnetic field properties lead to contraction of the ISCO to $\bar{r}_{\text{ISCO}} = 4.891$, giving rise to a greater efficiency $\eta \approx 0.0687$. In an anti-de Sitter background, the negative cosmological constant shifts the potential barrier. Therefore, the setup Magnetic AdS features expansion of the inner boundary to $\bar{r}_{\text{ISCO}} = 5.565$ and attains the minimum efficiency value of $\eta \approx 0.0414$.

\begin{table}
\centering
\caption{Radiative efficiency $\eta = 1 - \bar{E}_{\text{ISCO}}$ for magnetic black hole case.}
\label{tab:efficiency-magnetic}
\begin{tabular}{lccc}
\hline
Model & $\bar{r}_{\text{ISCO}}$ & $\bar{E}_{ISCO}$ & $\eta$ \\ \hline
Schwarzschild         & 6.000 & 0.943 & 0.0570 \\
RN ($q=0.8,p=0.1$)    & 5.402 & 0.937 & 0.0629 \\
RN ($q=0.3,p=0.1$)AdS & 5.455 & 0.957 & 0.0430 \\
Magnetic 1                & 5.985 & 0.943 & 0.0573 \\
Magnetic 2                & 4.891 & 0.931 & 0.0687 \\
Magnetic AdS              & 5.565 & 0.959 & 0.0414 \\ \hline
\end{tabular}
\end{table}

\begin{figure*}
\centering
\includegraphics[width=0.9\textwidth]{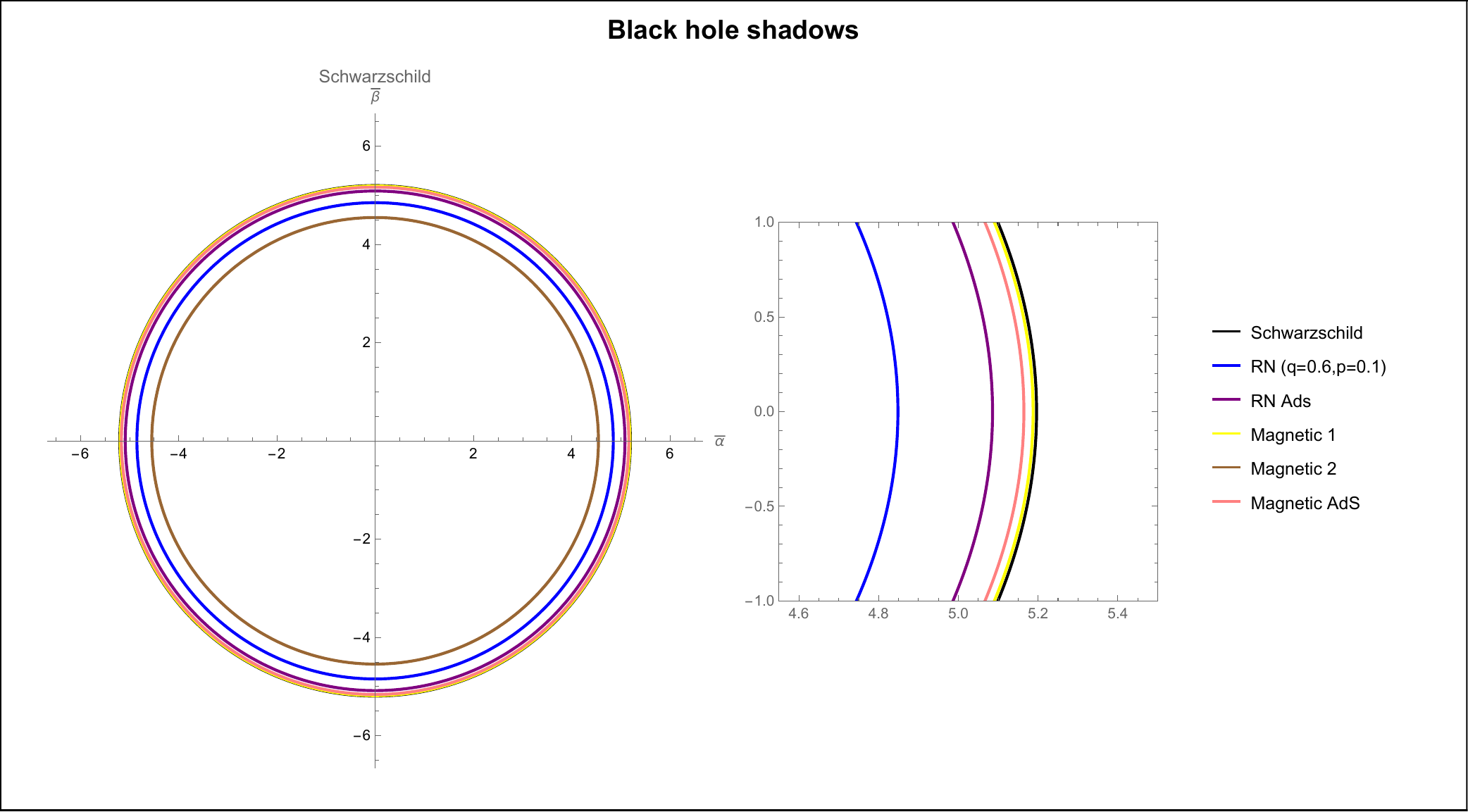}
\caption{Celestial plane shadows (circles) for the different black hole models. The radius of each circle is the critical impact parameter $\bar{b}_{\text{crit}}$. The Schwarzschild shadow is shown as a reference.}
\label{fig:shadowsMag}
\end{figure*}

Through Fig. (\ref{fig:shadowsMag}) and Table~\ref{tab:shadows-magnetic}, we analyze in detail the physical characteristics of the photon sphere radius $\bar{r}_{\text{ph}}$ and the corresponding critical impact parameter $\bar{b}_{\text{crit}}$ under pure non-linear magnetic fields. Based on the geometric boundary detection, we observe that the presence of magnetic field parameters causes a significant shrinking of the optical surfaces with respect to the usual Schwarzschild bound value $(\bar{r}_{\text{ph}} = 3.000, \bar{b}_{\text{crit}} \approx 5.196)$. For instance, the Magnetic 1 profile tracks closely to the vacuum baseline, registering a shadow radius of $\bar{b}_{\text{crit}} \approx 5.187$. 

Nonetheless, this reduction becomes very obvious in the Magnetic 2 case, where the strong non-linear effect reduces the radius of the photon sphere to $\bar{r}_{\text{ph}} = 2.485$ and moves the critical shadow area closer to $\bar{b}_{\text{crit}} = 4.546$. This happens because the non-linear magnetic terms add attractive stress-tensor effects and change the background curvature in the core region. By projecting the paths on the observer plane through the concentric circles of Fig. (\ref{fig:shadowsMag}), one can clearly see the difference between these models and understand how the dark area size on the sky is strongly influenced by the magnetic coupling effect.

\begin{table}
\centering
\caption{Photon sphere radius $\bar{r}_{\text{ph}}$ and shadow radius $b_{\text{crit}}/M$ for the magnetic black holes.}
\label{tab:shadows-magnetic}
\begin{tabular}{lcc}
\hline
Model & $\bar{r}_{\text{ph}}$ & $\bar{b}_{\text{crit}}$ \\ \hline
Schwarzschild & 3.000 & 5.196 \\
RN ($q=0.8,p=0.1$) & 2.729 & 4.848 \\
RN ($q=0.3,p=0.1$)AdS & 2.932 & 5.086 \\
Magnetic 1 & 2.993 & 5.187 \\
Magnetic 2 & 2.485 & 4.546 \\
Magnetic AdS & 2.993 & 5.164 \\ \hline
\end{tabular}
\end{table}

\begin{figure}
\centering
\includegraphics[width=0.5\textwidth]{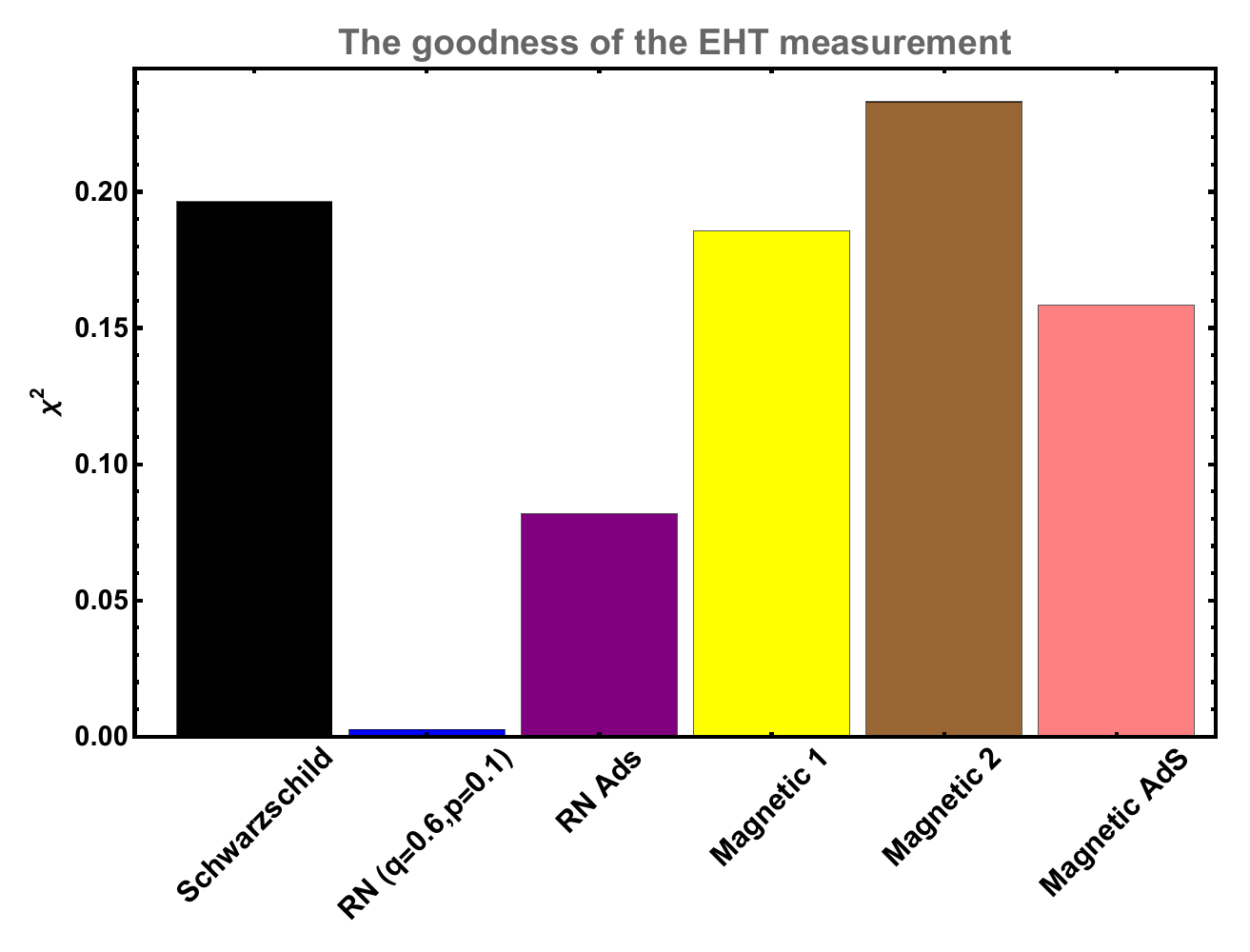}
\caption{The $\chi^{2}$ of angular diameters of the black hole shadow for the different models.}
\label{X2Mag}
\end{figure}

\begin{figure}
\centering
\includegraphics[width=0.5\textwidth]{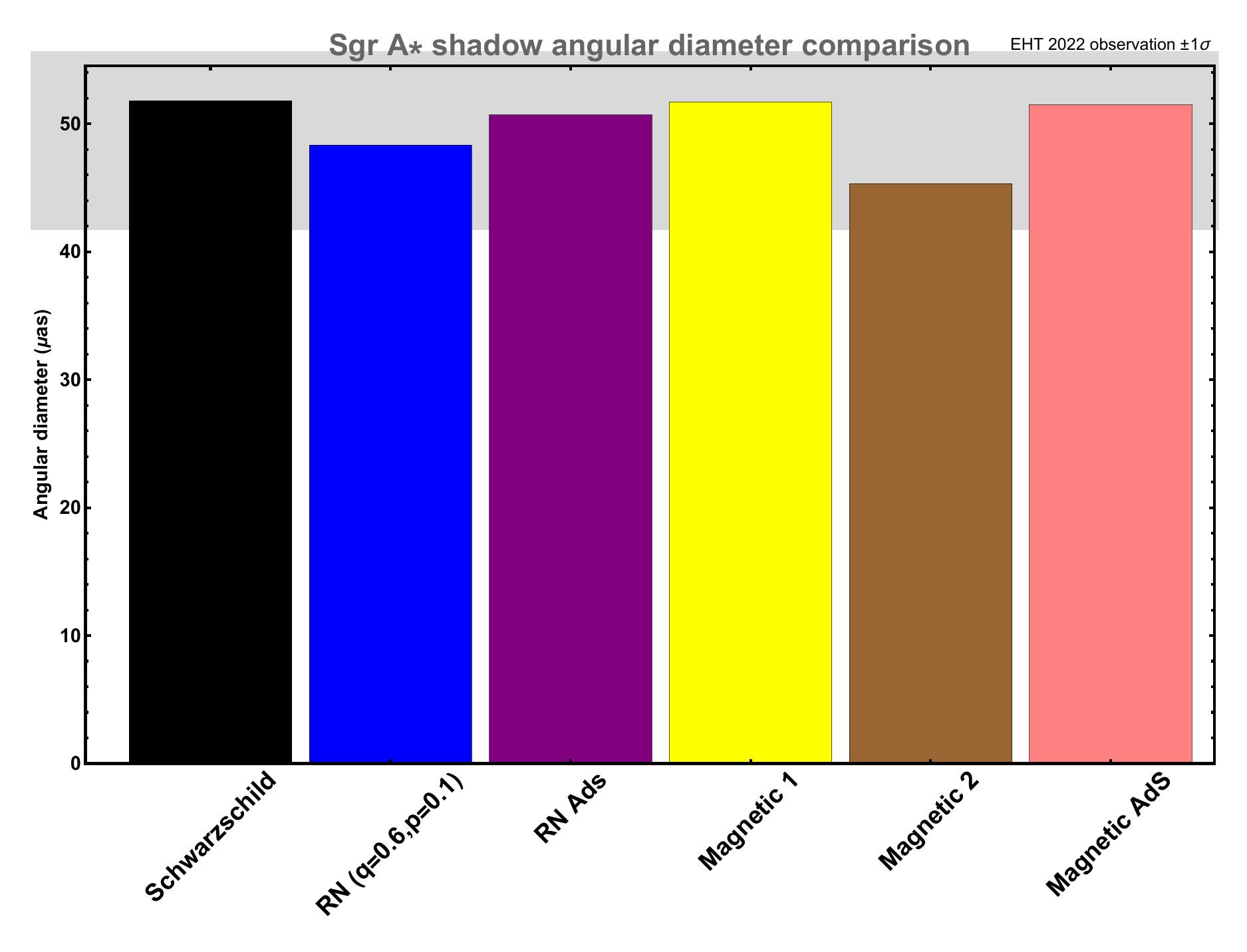}
\caption{Angular diameters of the black hole shadow for the different models, compared with the EHT observation of Sgr A* (dashed line) and its $1\sigma$ uncertainty (gray band). Models with bars intersecting the gray band are consistent with the data at the $1\sigma$ level.}
\label{fig:sgrA_comparisonMag}
\end{figure}

In order to carry out an observational test of our purely magnetic framework, we collect the physical values of angular diameter ($\theta$) and their corresponding statistical values of $\chi^2$ measured according to the Sgr A* scaling configuration in Table~\ref{tab:sgra_comparison-magnetic}. The resulting probability distribution is demonstrated in Figs.~(\ref{X2Mag}) and (\ref{fig:sgrA_comparisonMag}). The empirical value of the horizon ($\theta_{\text{obs}} = 48.7 \pm 7.0\,\mu\text{as}$) measured by the Event Horizon Telescope provides us with a physical constraint.

As seen from the probability graphs, all the proposed variations of magnetic models provide us with apparent diameters lying in between the $1\sigma$ and $2\sigma$ confidence intervals. The reference standard of the Schwarzschild metric lies slightly above the central line, having $\chi^2 \approx 0.1963$. Our variations of the Magnetic 1 and Magnetic AdS models follow suit, with $\chi^2$ values of $0.1855$ and $0.1583$, respectively. What is more, the small diameter of the shadow boundary of the Magnetic 2 model equals $\theta \approx 45.32\,\mu\text{as}$ ($\chi^2 \approx 0.2330$). Thus, this is the lowest observational bound for our framework. The overall matching presented in Fig.~(\ref{fig:sgrA_comparisonMag}) shows that non-linear magnetic fields demonstrate high statistical compatibility with current measurements of the horizon.

\begin{table}
\centering
\caption{Angular diameter of the black hole shadow for magnetic black holes, compared with the EHT measurement for Sgr A*. The $\chi^2$ value indicates the goodness of fit.}
\label{tab:sgra_comparison-magnetic}
\begin{tabular}{lccc}
\hline
Model & $\bar{b}_{\text{crit}}$ & $\theta$ ($\mu$as) & $\chi^2$ \\ \hline
Schwarzschild         & 5.196 & 51.80 & 0.1963 \\
RN ($q=0.8,p=0.1$)    & 4.849 & 48.34 & 0.0027 \\
RN ($q=0.3,p=0.1$)AdS & 5.086 & 50.70 & 0.0817 \\
Magnetic 1                & 5.187 & 51.72 & 0.1855 \\
Magnetic 2                & 4.546 & 45.32 & 0.2330 \\
Magnetic AdS              & 5.164 & 51.49 & 0.1583 \\ \hline
\end{tabular}
\end{table}

\section{Conclusions}\label{sec:6}

In this work, we systematically investigated the observational signatures of static, spherically symmetric dyonic black holes in non-linear electrodynamics, focusing on thin accretion disk properties and the resulting electromagnetic radiation. It should be mentioned that our primary objective was to understand how the modifications to the gravitational potential arising from the NLED framework, specifically the asymmetric treatment of electric and magnetic charges, manifest in astrophysical observables that can be tested against current and future observations.

In this part of our study, we considered the most general metric of a dyonic black hole extracted from the NLED action, depending on six parameters: mass, cosmological constant, electric and magnetic charge, and NLED coupling constants. Due to the complexity of the problem, the electrostatic field $F_{rt}$ depends on a third-degree polynomial and could be analytically solved only for the case $q = 0$. In this case, we obtained an analytical solution that shows the most important modification of the metric function $f(r)$ due to the $1/r^6$ term. The most intriguing fact about this term is the increase of its contribution with the distance to the event horizon; therefore, it will make essential contributions to the deviation of the black hole solution from the standard Schwarzschild or Reissner-Nordstr\"om cases.

Following our study of particle dynamics, we derived the equations of motion and the effective potential for test particles in the equatorial plane. As a result, the effective potential barriers are strongly affected by the electric and magnetic charges, as well as the NLED coupling constants. It is also clear that the innermost stable circular orbit moves inward and its radius decreases as the charge parameters increase. This effect will strongly impact the physics of accretion disks.

Finally, we considered the kinematics of circular orbits for particles around a black hole, focusing on the angular velocity, specific energy, and specific angular momentum. We have shown that the angular velocity of particles increases monotonically as they approach the black hole and approaches unity near the horizon. The specific energy and specific angular momentum show three-stage behavior, decreasing as they approach the black hole and increasing near the event horizon. We would like to underline the resemblance in behavior between the RN AdS and NLED AdS cases.

For the thermodynamic characteristics of the accretion disc, we calculated the energy flux and effective temperature using the conventional Page-Thorne model. Our results show that the flux rises slowly for particles falling into the black hole, reaches a maximum, and then drops suddenly as they approach the event horizon. As expected, black holes with the Schwarzschild metric show the minimal flux values, whereas the RN AdS configuration yields the maximal values, with the NLED metrics in between. Moreover, we observe that electric and magnetic charges shift the peak of thermal radiation toward smaller radii and considerably increase its value compared with the Schwarzschild vacuum solution. The reason for this phenomenon is that the closer the position of the ISCO, the greater is the energy released by the accretion disc.

Additionally, we studied the radiative efficiency of the accretion process, defined as the ratio of rest-mass energy converted into radiation before the matter falls into the black hole. Our estimates show that the efficiency of the Schwarzschild black hole is roughly $5.72\%$, and the deviations in the efficiencies of the NLED models do not exceed $10\%$ relative to this value. Specifically, the configuration with NLED 1, where $a=0.00037, b=0.028, q=0.8, p=0.1$, reaches the maximum efficiency of $6.90\%$. The minimum efficiency belongs to the NLED AdS model and equals $4.20\%$. These deviations make it an attractive tool for verifying NLED models using X-ray observations of quasars and active galactic nuclei.

Beyond thermal characteristics, we studied the optical properties of the black hole spacetime by determining the photon sphere radius and the critical impact parameter that defines the size of the black hole shadow. We found that, for all charged cases, the photon sphere radius is less than $3$, characteristic of the Schwarzschild black hole, and it decreases further with increasing charge. The shadows of NLED spacetimes have sizes ranging between about $4.534$ (NLED 1) and $5.159$ (NLED 2), while the Schwarzschild spacetime has a shadow size of $5.196$.

It is quite clear that the most important finding of our study is the comparison of the theoretical shadow angular diameters with EHT observations of Sagittarius A*. We used the black hole's physical mass and distance to convert our dimensionless shadow radii to angular diameters and performed a $\chi^2$ goodness-of-fit test. It must be noted that all of the NLED models proposed in this study have shadow sizes that fall under the $1\sigma$ and $2\sigma$ ranges of the EHT measurements. The shadow angular diameter for Schwarzschild is $51.80\,\mu\text{as}$ with $\chi^2 \approx 0.196$, while the angular diameters of the NLED models vary between $45.20\,\mu\text{as}$ and $51.43\,\mu\text{as}$, with $\chi^2 < 0.26$.

In the purely magnetic case, we found that the thin accretion disk properties are similar to those in the dyonic case. The Magnetic 2 configuration has the minimal ISCO radius of $4.891$ and the maximal efficiency of $\sim 6.87\%$. In the Magnetic 2 case, the shadow radius is $4.546$, and the angular diameter of the shadow is $45.32 \mu$as with $\chi^2 \approx 0.233$, which is the lowest observational constraint for the considered magnetic configurations. Furthermore, the shadow radius of the Magnetic 1 and Magnetic AdS models is close to the Schwarzschild shadow radius, with the $\chi^2$ values of $0.186$ and $0.158$, respectively. The negative cosmological constant in the magnetic AdS model increases the inner radius of the accretion disk and decreases the system's efficiency. Therefore, it is demonstrated that there is competition between the NLED effects and the cosmological constant in such cases.

In summary, our results prove that the thin accretion disk around dyonic and purely magnetic NLED black holes displays distinct observational signatures that could be used in future observations. The modified gravitational potential allows us to detect changes in the ISCO radius, thermal spectrum, radiative efficiency, photon sphere, and black hole shadow. The good agreement between the considered NLED models and EHT observations of Sgr A* confirms the viability of the non-linear electrodynamics framework for describing the strong-field regime of black holes. Future precise observations from the next-generation EHT and other interferometric telescopes will allow us to test different NLED models and place restrictions on the NLED coupling constants $a$ and $b$. Therefore, we provide templates for future observations and probe the gravity-electromagnetism interaction in the strong-field regime.

\section*{Acknowledgements}
This work is supported by the National Natural Science Foundation of China (No. 12675133).



\end{document}